\documentclass[preprint,authoryear,12pt]{elsarticle}

\usepackage{amssymb}
\usepackage{amsmath}
\usepackage{multirow}
\usepackage{amsfonts}
\usepackage{mathrsfs}
\usepackage{xcolor}
\usepackage{booktabs}
\usepackage{algorithm}
\usepackage{algorithmicx}
\usepackage{algpseudocode}
\usepackage{bm}
\usepackage{hyperref}
\usepackage{siunitx}
\usepackage{subcaption}

\begin{document}

\begin{frontmatter}

\title{Portable surrogate-free 4D MRI from standard fast multi-slice 2D MRI via implicit neural representations\tnoteref{t1}}
\tnotetext[t1]{Project page: \href{https://ttlmh.github.io/simple-4d/}{https://ttlmh.github.io/simple-4d/}}

\author[psi,d-phys]{Muheng Li}
\author[psi]{Xinyang Wu}
\author[psi,d-infk]{Xia Li}
\author[unibasp,uniba,ibt]{Orso Pusterla}
\author[uniba]{Philippe C. Cattin}
\author[psi]{Sairos Safai}
\author[psi,d-phys]{Antony J. Lomax}
\author[psi]{Ye Zhang\corref{cor1}}
\ead{ye.zhang@psi.ch}
\cortext[cor1]{Corresponding author.}
\affiliation[psi]{organization={Center for Proton Therapy, Paul Scherrer Institut},
            city={Villigen},
            country={Switzerland}}
\affiliation[d-phys]{organization={Department of Physics, ETH Zurich},
            city={Zurich},
            country={Switzerland}}
\affiliation[d-infk]{organization={Department of Computer Science, ETH Zurich},
            city={Zurich},
            country={Switzerland}}
\affiliation[unibasp]{organization={Department of Radiology, Division of Radiological Physics, University Hospital Basel},
            city={Basel},
            country={Switzerland}}
\affiliation[uniba]{organization={Department of Biomedical Engineering, University of Basel},
            city={Allschwil},
            country={Switzerland}}
\affiliation[ibt]{organization={Institute for Biomedical Engineering, University and ETH Zurich},
            city={Zurich},
            country={Switzerland}}

\begin{abstract}
Four-dimensional MRI (4D MRI) characterizes respiratory organ motion, yet
existing reconstruction pipelines are tightly coupled to specific acquisition
platforms (e.g., non-Cartesian trajectories with self-gating, vendor-specific
navigators, or external respiratory hardware), limiting broad adoption across
diverse clinical and research settings, including low-field, open-bore, and
non-supine imaging. We present \textbf{SIMPLE-4D} (\textbf{S}urrogate-free,
\textbf{IM}plicit, \textbf{P}ortab\textbf{LE} 4D MRI), a software-first
portable workflow that operates entirely on reconstructed slices from
standard fast multi-slice 2D MRI and requires no pulse-sequence modification,
no non-Cartesian trajectory, no navigator, and no external hardware.
SIMPLE-4D combines an acquisition-agnostic front end consuming standard 2D
protocols, a surrogate-free variational motion encoder
that extracts a compact motion code directly from each 2D slice, and a
physics-aware continuous spatio-temporal reconstruction based on a
hash-encoded implicit neural representation (INR) with a SIREN deformation
network producing bidirectional cycle-consistent DVFs and motion-dependent
Gauss--Legendre thick-slice quadrature. Bidirectionality yields a complete
inter-frame motion model by composition, supporting downstream tasks such
as dose accumulation without retraining. We validate the identical pipeline
on two contrasting datasets: a \SI{1.5}{\tesla} clinical bSSFP dataset
(5 volunteers, 3 sessions each) and a \SI{0.5}{\tesla} open-bore HASTE
dataset (5 volunteers, supine and upright). To our knowledge, this is the
first per-frame 4D volumetric respiratory MRI reconstruction on a
weight-bearing upright open-bore low-field scanner from reconstructed 2D
Cartesian slices alone. On low-field data the INR template additionally
acts as an implicit denoiser, yielding $+132\%$ SNR. Systematic ablations
isolate each component's contribution.
\end{abstract}


\end{frontmatter}


\section{Introduction}\label{sec:intro}

Respiratory motion is a dominant source of geometric uncertainty in
radiation therapy of thoracic and abdominal tumours, causing patient-specific
target displacement~\citep{seppenwoolde2002}, degrading dose
conformality~\citep{colvill2016}, and varying between and within fractions in
ways planning 4D-CT cannot capture~\citep{ge2013,dhont2018}.
Four-dimensional MRI (4D MRI) has become the preferred modality for
characterising this motion, with superior soft-tissue contrast, no ionising
radiation, and potential for real-time volumetric imaging on MR-guided
linear accelerator (MR-Linac)
platforms~\citep{paganelli2018,stemkens2018}.

Despite two decades of methodological progress, 4D MRI is not as broadly
adopted as its imaging strengths would predict. The root cause is not image
quality but \emph{portability}: dominant pipelines are tightly coupled to
specific scanner--sequence combinations. We organise these couplings into
seven portability axes:
\begin{enumerate}
    \item \textbf{Acquisition-trajectory dependence.} Golden-angle radial,
    stack-of-stars, or 3D non-Cartesian trajectories that oversample the
    k-space centre for
    self-gating~\citep{feng2016xdgrasp,feng2020graspro,ong2020,han2017rock}
    require well-calibrated non-Cartesian
    sampling~\citep{keall2022mrgrt,thorwarth2021}.

    \item \textbf{Pulse-sequence / vendor dependence.} Interleaved
    navigators~\citep{vonsiebenthal2007,cai2011} and most self-gated
    acquisitions need pulse-sequence modifications in vendor-specific
    environments most clinical sites cannot
    access~\citep{layton2017pulseq,karakuzu2022vendorneutral}.

    \item \textbf{External respiratory hardware.} Bellows and optical
    trackers~\citep{li2017bellows} add setup time, correlate imperfectly
    with internal organ motion~\citep{seppenwoolde2002,dhont2018}, and may
    be incompatible with MR-Linac
    geometries~\citep{campbell2023lowfield}.

    \item \textbf{1D respiratory surrogate.} Reconstruction is typically
    driven by a \emph{scalar} signal (navigator-derived diaphragm position,
    bellows, or k-space-centre self-gating), which cannot represent
    hysteresis, differential motion between organs, or non-periodic
    dynamics~\citep{seppenwoolde2002,ge2013,dhont2018}.

    \item \textbf{Acquisition-geometry dependence.} Several MR-Linac methods
    extract motion only from one to three fixed slice geometries, such as
    orthogonal sagittal/coronal cines~\citep{rabe2021porcine,xiong2023continuous,jassar2023orthogonal}
    or a dedicated navigator slice~\citep{vonsiebenthal2007,wu2025cpt4dmr},
    and cannot be deployed on standard multi-slice 2D protocols whose
    slice prescription is dictated by the clinical workflow.

    \item \textbf{Reference- or prior-scan dependence.} Many model-based
    methods require a separately acquired anatomical prior such as a 3D
    breath-hold reference~\citep{huttinga2020}, a pre-treatment 4D
    scan~\citep{shao2025dremeMR}, or a pre-built signature
    dictionary~\citep{feng2023mrsigma}; breath-hold references in particular
    are problematic for cancer patients, who often cannot sustain a
    reproducible breath-hold of sufficient duration.

    \item \textbf{Per-patient optimisation cost.} Many INR-based methods
    (including ours) require per-patient training (minutes to hours);
    population-trained networks amortise this cost but introduce a
    generalisation penalty.
\end{enumerate}

The proposed framework removes coupling along axes~1--6. Because it
operates in image space, residual sensitivity to contrast and receive-coil
configuration remains.

\begin{figure*}[htbp]
    \centering
    \includegraphics[width=0.98\textwidth]{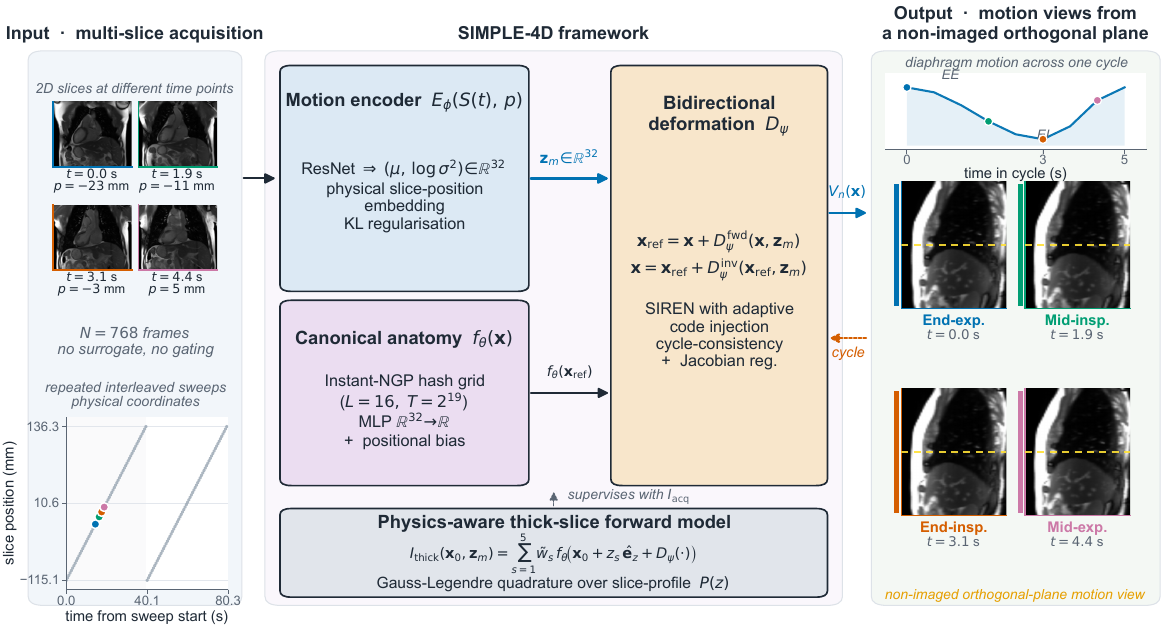}
    \caption{Overview of the proposed SIMPLE-4D workflow. \textbf{Left:} time-ordered multi-slice 2D Cartesian acquisitions at varying slice positions. \textbf{Centre:} variational motion encoder, hash-encoded canonical anatomy, bidirectional deformation network, and thick-slice forward model. \textbf{Right:} reconstructed per-frame 3D volumes across four respiratory phases.}
    \label{fig:framework}
\end{figure*}

\paragraph{A software-first alternative}
We argue that the path to portable 4D MRI is not another custom sequence,
but a \emph{software-first} post-processing workflow that sits on top of
the acquisition every clinical scanner already supports: standard fast
multi-slice 2D MRI. We refer to the proposed workflow as \textbf{SIMPLE-4D}
(\textbf{S}urrogate-free, \textbf{IM}plicit, \textbf{P}ortab\textbf{LE}
4D MRI). It takes only a time-ordered sequence of reconstructed 2D slices
with known spatial positions, and combines (Fig.~\ref{fig:framework}):
(i)~an \emph{acquisition-agnostic front end} consuming any standard
2D Cartesian multi-slice dynamic series in image space;
(ii)~an \emph{image-domain surrogate-free motion cue}, a variational
encoder mapping each slice and its slice-position embedding to a
32-dimensional motion code that replaces scalar surrogates; and
(iii)~a \emph{continuous spatio-temporal reconstruction} pairing a
hash-encoded INR canonical anatomy~\citep{muller2022instantngp} with a
SIREN~\citep{sitzmann2020siren} conditional deformation network that
predicts bidirectional cycle-consistent DVFs~\citep{wolterink2022idir}, and
a physics-aware Gauss--Legendre thick-slice forward model accounting for
the finite slice excitation profile. The output is a temporally dense
per-frame 3D volume sequence without phase binning or periodicity
assumptions. (In addition, vendor-neutral acquisition frameworks such as Pulseq~\citep{layton2017pulseq} can be composed with this reconstruction for fully vendor-neutral 4D MRI.)

\paragraph{Cross-platform validation}
We validate the same unmodified pipeline on a \SI{1.5}{\tesla} clinical bSSFP dataset and a \SI{0.5}{\tesla} open-bore HASTE dataset (supine and upright), demonstrating robustness across field strength, scanner geometry, pulse sequence, and patient positioning.

\paragraph{Contributions}
\begin{enumerate}
    \item \textbf{SIMPLE-4D}, a software-first, portable 4D MRI workflow
    operating entirely on reconstructed slices from standard fast
    multi-slice 2D MRI, removing six of the seven acquisition couplings
    above simultaneously (axes~1--6).

    \item A \textbf{surrogate-free variational motion encoder} extracting a
    compact motion code from each 2D slice and its slice-position
    embedding, learned end-to-end with the INR rather than as a separate
    preprocessing step, in contrast to prior image-domain manifold
    methods~\citep{wachinger2012manifold,chen2017manifold,romaguera2022population,zou2022manifold}.

    \item A \textbf{physics-aware INR reconstruction} combining a
    hash-encoded canonical anatomy, a SIREN deformation network with
    adaptive multi-scale motion-code injection, bidirectional cycle-consistent
    DVFs, and motion-dependent Gauss--Legendre thick-slice supervision.

    \item \textbf{Cross-platform validation} on \SI{1.5}{\tesla} bSSFP and
    \SI{0.5}{\tesla} open-bore HASTE in supine and upright positions with
    one unmodified pipeline, including, to our knowledge, the \emph{first}
    per-frame 4D volumetric respiratory MRI reconstruction on a
    \emph{weight-bearing upright open-bore low-field} scanner from
    reconstructed 2D Cartesian slices alone. As an additional benefit,
    the INR template provides \textbf{implicit denoising} on low-field
    data ($+132\%$ SNR).

    \item A \textbf{systematic ablation} of INR backbone, motion encoder,
    deformation conditioning, bidirectional cycle-consistent DVF, and
    thick-slice supervision, providing architectural guidance for INR-based
    dynamic medical imaging more broadly.
\end{enumerate}

\section{Related work}\label{sec:related}

We organise prior work around the seven portability axes identified in
Section~\ref{sec:intro}. Table~\ref{tab:method_comparison} summarises where
existing methods sit along axes~1--6 (axis~7, per-patient optimisation cost,
is a property of the reconstruction family rather than the acquisition
platform and is discussed in Section~\ref{sec:discussion}).

\begin{table}[htbp]
\centering
\caption{Comparison of 4D MRI reconstruction approaches along the six
acquisition-level portability axes of Section~\ref{sec:intro}.
\textbf{Cart}: Cartesian-only compatible. \textbf{NoSeqMod}: no
vendor-specific pulse-sequence modification. \textbf{NoExtHW}: no external
respiratory hardware. \textbf{NoSurr}: no 1D respiratory surrogate.
\textbf{AnyGeom}: no specific slice-geometry dependence. \textbf{NoRef}:
no pre-acquired reference or dictionary scan. \textbf{P/F}: per-frame
volumetric output. \textbf{Low-F}: validated at $\le \SI{0.5}{\tesla}$.
Method families with matching axis profiles are merged into single rows;
\citet{romaguera2022population} is omitted because its input modality is
2D ultrasound (discussed in Section~\ref{sec:related_imagesurrogate}).}
\label{tab:method_comparison}
\scriptsize
\setlength{\tabcolsep}{2pt}
\renewcommand{\arraystretch}{0.95}
\resizebox{0.97\textwidth}{!}{%
\begin{tabular}{@{}l p{1.55cm} p{2.0cm} ccccccccc@{}}
\toprule
Method & Surrogate & Acquisition & Cart & NoSeqMod & NoExtHW & NoSurr & AnyGeom & NoRef & P/F & Low-F \\
\midrule
Phase-sorted~\citep{vonsiebenthal2007}                                      & Navigator   & Multi-slice 2D         & \checkmark & \texttimes & \checkmark & \checkmark & \texttimes & \checkmark & \checkmark & \texttimes \\
Image-scalar~\citep{cai2011,liu2017scdpoi,tong2021flux}                     & Image-1D    & Multi-slice 2D         & \checkmark & \checkmark & \checkmark & \texttimes & \texttimes & \checkmark & \texttimes & \texttimes \\
Manifold image surrog.~\citep{wachinger2012manifold}                        & Image-ND    & Multi-slice 2D         & \checkmark & \texttimes & \checkmark & \texttimes & \texttimes & \checkmark & \texttimes & \texttimes \\
Variational latent manifold~\citep{zou2022manifold}                         & Image-ND    & Multi-slice 2D spiral  & \texttimes & \texttimes & \checkmark & \checkmark & \checkmark & \checkmark & \checkmark & \texttimes \\
Manifold k-space surrog.~\citep{chen2017manifold}                          & k-space-ND  & Golden-angle radial    & \texttimes & \texttimes & \checkmark & \checkmark & \checkmark & \checkmark & \checkmark & \texttimes \\
XD-GRASP~\citep{feng2016xdgrasp}                                            & Self-gated  & Golden-angle radial    & \texttimes & \texttimes & \checkmark & \texttimes & \checkmark & \checkmark & \texttimes & \texttimes \\
Extreme MRI~\citep{ong2020}                                                 & Self-gated  & 3D non-Cart.           & \texttimes & \texttimes & \checkmark & \checkmark & \checkmark & \checkmark & \checkmark & \texttimes \\
MR-MOTUS~\citep{huttinga2020}                                              & None        & 3D Cart.    & \checkmark & \texttimes & \checkmark & \checkmark & \checkmark & \texttimes & \checkmark & \texttimes \\
Low-rank MR-MOTUS~\citep{huttinga2021lowrank,huttinga2022tmotus}            & Self-gated  & Golden-angle radial    & \texttimes & \texttimes & \checkmark & \texttimes & \checkmark & \checkmark & \checkmark & \texttimes \\
Eiben et al.~\citep{eiben2024}                                              & Image-1D (internal) & Multi-slice 2D & \checkmark & \texttimes & \checkmark & \checkmark & \texttimes & \checkmark & \checkmark & \texttimes \\
ICoNIK family~\citep{spieker2024,spieker2024selfsup}                        & Self-gated  & Radial stack-of-stars  & \texttimes & \texttimes & \checkmark & \texttimes & \checkmark & \checkmark & \texttimes & \texttimes \\
MODEST~\citep{terpstra2023modest}                                           & Self-gated  & Golden-angle radial    & \texttimes & \texttimes & \checkmark & \texttimes & \checkmark & \checkmark & \texttimes & \texttimes \\
Movienet~\citep{murray2024movienet}                                         & Self-gated  & 3D radial stack        & \texttimes & \texttimes & \checkmark & \texttimes & \checkmark & \checkmark & \texttimes & \texttimes \\
Rabe / Xiong~\citep{rabe2021porcine,xiong2023continuous}                    & Image-based & Orth.\ 2D cine (0.35T) & \checkmark & \checkmark & \checkmark & \checkmark & \texttimes & \texttimes & \checkmark & \checkmark \\
CAPTURE~\citep{chen2023capture}                                             & Self-gated  & Rapid radial (0.35T)   & \texttimes & \texttimes & \checkmark & \texttimes & \checkmark & \checkmark & \texttimes & \checkmark \\
MRSIGMA~\citep{feng2023mrsigma}                                             & Self-gated  & Radial + dict.\ (1.5T) & \texttimes & \texttimes & \checkmark & \texttimes & \checkmark & \texttimes & \checkmark & \texttimes \\
CPT-4DMR~\citep{wu2025cpt4dmr}                                              & Navigator   & Multi-slice 2D         & \checkmark & \texttimes & \checkmark & \texttimes & \texttimes & \checkmark & \checkmark & \texttimes \\
STINR-MR~\citep{shao2024stinr}                                              & Self-gated  & 3D radial              & \texttimes & \texttimes & \checkmark & \texttimes & \checkmark & \checkmark & \checkmark & \texttimes \\
DREME-MR~\citep{shao2025dremeMR}                                            & Learned latent & 3D radial           & \texttimes & \texttimes & \checkmark & \checkmark & \checkmark & \texttimes & \checkmark & \texttimes \\
Kunz et al.~\citep{kunz2024cardiac}                                         & None        & Cartesian multi-shot   & \checkmark & \checkmark & \checkmark & \checkmark & \checkmark & \checkmark & \texttimes & \texttimes \\
\midrule
\textbf{SIMPLE-4D (Proposed)} & \textbf{None} & \textbf{Any 2D Cart.\ m.-s.\ dyn.} & \checkmark & \checkmark & \checkmark & \checkmark & \checkmark & \checkmark & \checkmark & \checkmark \\
\bottomrule
\end{tabular}}
\end{table}

\subsection{Retrospective phase-sorted 4D MRI}\label{sec:related_sorting}

The established pipeline for 4D MRI acquires 2D slices over many breathing
cycles and bins them into a small number of discrete respiratory phases
using a respiratory surrogate, followed by inter-phase
registration~\citep{vonsiebenthal2007,cai2011,li2017bellows}. The surrogate
may be an \emph{internal} signal derived from interleaved navigator slices
acquired within the same sequence~\citep{vonsiebenthal2007}, or an
\emph{external} signal such as a respiratory bellows or optical
tracker~\citep{li2017bellows}; we use ``external surrogate'' throughout in
the latter, hardware sense and treat navigator-derived signals as internal
image-based surrogates. Subsequent
extensions improve robustness to irregular breathing through
amplitude-binning outlier rejection~\citep{vankesteren2019} and validate
clinical applicability against 4D-CT~\citep{peteani2024}. Phase sorting is
clinically mature but inherits all four acquisition couplings: it requires
a surrogate, assumes respiratory regularity, caps temporal resolution at a
small number of phases, and averages away breath-to-breath variability. The
output of phase binning is therefore not a true per-frame time series but a
single representative 4D dataset of one \emph{average} breathing cycle, in
which inter-cycle variability and non-periodic dynamics have been collapsed.

\subsection{Non-Cartesian self-gated motion-resolved imaging}\label{sec:related_kspace}

Continuous non-Cartesian acquisition with k-space-centre oversampling
enables self-gated motion-resolved imaging (e.g., XD-GRASP~\citep{feng2016xdgrasp},
Extreme MRI~\citep{ong2020}, MR Multitasking~\citep{christodoulou2018multitasking}).
These methods achieve excellent temporal resolution but inherently couple
to non-Cartesian trajectories.

\subsection{Model-based joint reconstruction and motion estimation}\label{sec:related_model}

MR-MOTUS~\citep{huttinga2020} factorises the space--time motion field via
a joint inverse problem from minimal k-space data using a pre-acquired
reference image~\citep{huttinga2021lowrank,huttinga2022tmotus}.
McClelland et
al.~\citep{mcclelland2013motionmodels} review respiratory motion models more
broadly. These approaches offer strong inverse-problem formulations but
typically retain coupling through specialised acquisitions, pre-acquired
reference scans, or external surrogate signals.

\subsection{Image-based respiratory motion representations}\label{sec:related_imagesurrogate}

A first line of work replaces external hardware with image-derived
\emph{scalar} surrogates extracted from a dedicated 2D plane: the
time-varying torso cross-sectional area~\citep{cai2011,vankesteren2019}, or
the geometric intersection of sagittal, coronal, and diaphragm
planes~\citep{liu2017scdpoi}. These methods take an important step: they
recognise that image content already contains respiratory information, but
remain within the signal-driven paradigm: the motion representation is
a 1D scalar, a dedicated plane must be reserved for surrogate extraction,
and downstream reconstruction still uses discrete phase binning.

A second, more recent line replaces the scalar with a higher-dimensional
data-derived motion representation. Manifold learning produces
multi-dimensional respiratory embeddings either from full image
frames~\citep{wachinger2012manifold} or from the k-space data of radial
acquisitions~\citep{chen2017manifold}; PCA-based image-driven 3D abdominal
motion models~\citep{stemkens2016pca}, flux-based 4D-MRI from multi-slice
2D acquisitions~\citep{tong2021flux}, convolutional autoencoder latent
spaces~\citep{romaguera2021probabilistic,romaguera2022population}, and
variational manifolds learned from multi-slice dynamic
MRI~\citep{zou2022manifold} all establish that a data-derived,
high-dimensional motion representation can replace a 1D respiratory signal
for the purpose of organising motion-resolved data. Within this group,
\citet{romaguera2022population} additionally crosses modality by training a
conditional generative model from a 2D ultrasound surrogate to a 3D MRI
volume, which lies outside the multi-slice 2D MRI 4D reconstruction setting
considered in this paper.

Our work is conceptually aligned with this second line. We differ in three
specific ways: the motion code is extracted \emph{end-to-end jointly with
the reconstruction} (not as a separate preprocessing manifold), through a
\emph{variational single-frame} encoder applied to every acquired slice
together with its slice-position embedding, and is consumed by a
\emph{continuous} INR reconstruction that produces per-frame volumetric
output rather than phase-binned or 2D motion-estimated output.

\subsection{Implicit neural representations for dynamic MRI}\label{sec:related_inr}

INRs have recently been applied to medical image reconstruction.
SIREN~\citep{sitzmann2020siren} showed that periodic activations overcome the
spectral bias~\citep{rahaman2019spectral} of standard networks for
representing natural signals; Instant-NGP~\citep{muller2022instantngp}
introduced multi-resolution hash encoding for rapid convergence;
NeSVoR~\citep{xu2023nesvor} applies hash-encoded INRs with thick-slice
modelling to fetal-brain slice-to-volume reconstruction and provides the
methodological precedent for our forward model;
IDIR~\citep{wolterink2022idir} uses INRs for deformable registration. For
dynamic MRI, ICoNIK~\citep{spieker2024} and its self-supervised
extension~\citep{spieker2024selfsup} apply INRs in k-space for
respiratory-resolved abdominal MR using self-gating from radial
stack-of-stars; Kunz et al.~\citep{kunz2024cardiac} fit a Fourier-feature
INR directly to undersampled Cartesian k-space to reconstruct 2D
free-breathing \emph{cardiac} real-time cine, regularising jointly in space
and time. Kunz et al.\ is the closest method to ours on the
\emph{acquisition} side---it is Cartesian, surrogate-free, requires no
sequence modification, and uses no reference scan
(Table~\ref{tab:method_comparison})---but it addresses a different problem
and produces a different output: it reconstructs a 2D \emph{cardiac} image
time series rather than a per-frame 3D \emph{respiratory} volume, builds no
explicit respiratory motion model, and is not demonstrated at low field.
The present work differs in both problem and method, recovering per-frame
volumetric respiratory anatomy in image space from reconstructed
multi-slice 2D data through an explicit canonical-anatomy-plus-deformation
factorisation driven by an image-derived motion code, validated across
low-field and upright configurations.
The method most directly comparable to ours is
CPT-4DMR~\citep{wu2025cpt4dmr}, which jointly learns a Spatial Anatomy
Network and a Temporal Motion Network using periodic-activation INRs from
multi-slice 2D dynamic data, but is driven by a navigator-derived 1D
respiratory signal processed through a Transformer encoder. The present work differs by recovering respiratory state from image content via a variational encoder and validates on an out-of-distribution platform (0.5\,T open-bore HASTE, upright) with one unmodified pipeline.
STINR-MR~\citep{shao2024stinr} extends INR-based reconstruction to
respiratory 3D cine-MRI from golden-angle 3D radial data with PCA-derived
deformation bases from a prior 4D-MRI; its successor
DREME-MR~\citep{shao2025dremeMR} adds joint cardiorespiratory motion
estimation from a 3D radial pre-treatment scan via a frequency-guided
progressive training strategy.

Relative to this literature, the present work removes the two remaining
coupling axes that prior INR-based \emph{volumetric respiratory} 4D methods
retain: the non-Cartesian acquisition (ICoNIK, DREME-MR) and the 1D
respiratory surrogate (CPT-4DMR).

\subsection{4D MRI for MR-guided radiotherapy and at low field}\label{sec:related_rt}

MR-guided radiotherapy systems include ViewRay MRIdian
(\SI{0.35}{\tesla})~\citep{kluter2019} and Elekta Unity
(\SI{1.5}{\tesla})~\citep{raaymakers2017unity}. Both currently support
intra-fraction \emph{2D} cine motion
monitoring~\citep{thorwarth2021,keall2022mrgrt,rusu2025unity}; fully
volumetric 4D motion monitoring during beam delivery is available on
neither platform. Existing research approaches addressing this gap each
retain at least one acquisition coupling: orthogonal-plane geometry
constraints~\citep{rabe2021porcine,xiong2023continuous,jassar2023orthogonal},
self-gated radial trajectories couple
acquisition~\citep{chen2023capture,huttinga2023gp,shao2024stinr,shao2025dremeMR},
and dictionary-matching approaches require a pre-acquired 4D-MRI signature
scan~\citep{feng2023mrsigma}. 
Image-registration motion modelling for MR-guided lung
RT~\citep{eiben2024} drives the model with an internal image-derived
surrogate signal extracted from interleaved 2D surrogate slices, which
keeps a 1D respiratory surrogate (axis~4) in the training loop.
A pipeline that consumes only reconstructed 2D Cartesian
multi-slice images---with no special trajectory, no dedicated plane
geometry, and no reference scan---and from them reconstructs a
\emph{per-frame volumetric respiratory} 4D MRI remains, to our knowledge,
unaddressed. We make the qualifier explicit because the acquisition-level
portability properties alone are not unique to our work: Kunz
et al.~\citep{kunz2024cardiac} already satisfies all six acquisition axes of
Table~\ref{tab:method_comparison} (Cartesian, no sequence modification, no
external hardware, no surrogate, no fixed geometry, no reference scan), but
it reconstructs 2D real-time \emph{cardiac} cine and produces neither a
volumetric per-frame output (P/F) nor a respiratory motion model, and is not
demonstrated at low field (Low-F). What remains open is therefore the
combination of these six portability properties \emph{with} per-frame
volumetric respiratory output validated at low field---a combination
particularly valuable on low-field MR-Linacs, on open-bore systems without
navigator infrastructure, and on upright configurations.


\section{Methods}\label{sec:methods}

\subsection{Problem formulation}\label{sec:methods_formulation}

We formulate 4D MRI as per-frame slice-to-volume reconstruction. Given a time-ordered sequence of $N$ two-dimensional MRI slices $\{S_1, S_2, \ldots, S_N\}$, each acquired at a known spatial position $k_n \in \{0, \ldots, D-1\}$ and time index $n$, we seek to recover the complete 3D volume $V_n \in \mathbb{R}^{H \times W \times D}$ at each acquisition instant $n$. This is an ill-posed inverse problem: a single 2D slice cannot uniquely determine a 3D volume. We resolve this by decomposing the 4D reconstruction into per-frame deformations of an underlying canonical anatomy shared across time:
\begin{equation}
    V_n(\mathbf{x}) = f_\theta\!\left(\mathbf{x} + D_\psi\!\left(\mathbf{x},\, E_\phi(S_n,\, k_n)\right)\right)
    \label{eq:forward}
\end{equation}
where $f_\theta: [-1,1]^3 \to [0,1]$ is the canonical anatomy INR, $E_\phi: \mathbb{R}^{H \times W} \times \mathbb{Z} \to \mathbb{R}^{d_m}$ is the variational motion encoder that maps the single acquired slice $S_n$ and its slice-position index $k_n$ to a motion code $\mathbf{z}_m \in \mathbb{R}^{d_m}$, and $D_\psi: \mathbb{R}^3 \times \mathbb{R}^{d_m} \to \mathbb{R}^3$ is the conditional deformation network predicting displacement vector fields. The key insight is that the motion encoder $E_\phi$ replaces external surrogates: it discovers the respiratory state from image content alone.

  All spatial operations use normalised coordinates in $[-1,1]^3$
  placed at cell centres ($i \mapsto -1 + (2i+1)/S$ for a dimension of
  size $S$); the index ordering is (depth, width, height), corresponding
  to the slice-selection, left--right, and superior--inferior anatomical
  directions.

\subsection{Canonical anatomy network}\label{sec:methods_anatomy}

The canonical anatomy is encoded as a continuous function $f_\theta: [-1,1]^3 \to [0,1]$ using Instant-NGP~\citep{muller2022instantngp} with multi-resolution hash encoding (hyperparameters in \ref{app:hyperparams}). The explicit spatial data structure provides fast convergence, critical for short per-patient training budgets required clinically.

\subsection{Image-driven variational motion encoder}\label{sec:methods_encoder}

The motion encoder is the core innovation enabling surrogate-free reconstruction. It extracts a compact latent motion code $\mathbf{z}_m \in \mathbb{R}^{32}$ from a single 2D MRI slice together with its slice-position index, effectively replacing external respiratory signals with learned motion representations.

\subsubsection{Architecture}

A ResNet-style convolutional backbone with sinusoidal positional embedding of slice location produces a motion code. Architecture details (channel progression, MLP dimensions, dropout rates) are provided in \ref{app:hyperparams}.

\subsubsection{Slice-position conditioning}

Operating on the single acquired slice $S_n$ plus its slice-position index $k_n$ (rather than a temporal window of neighbouring frames) keeps the encoder strictly causal with respect to what has been acquired and removes the need to buffer or pre-align multiple frames at inference. The apparent ambiguity created by respiratory hysteresis (an identical diaphragm position can occur in both inhalation and exhalation) is resolved implicitly at the reconstruction level by the variational latent space: every slice is associated with a posterior $\mathcal{N}(\bm{\mu},\bm{\sigma}^2)$ whose geometry is shared across all slice positions, and the downstream INR plus deformation network only produces a consistent reconstruction when the code encodes a globally coherent respiratory state. The encoder ablation (Section~\ref{sec:results_main}) confirms that the resulting 32-dimensional embedding correlates strongly with independently tracked diaphragm motion without being exposed to that information during training.

\subsubsection{Variational inference}

Following the VAE framework~\citep{kingma2014vae}, the final MLP outputs mean $\bm{\mu}$ and log-variance $\log \bm{\sigma}^2$ vectors. The motion code is sampled via reparametrisation:
\begin{equation}
    \mathbf{z}_m = \bm{\mu} + \bm{\epsilon} \odot \exp(\log \bm{\sigma}^2 / 2), \quad \bm{\epsilon} \sim \mathcal{N}(\mathbf{0}, \mathbf{I})
    \label{eq:reparam}
\end{equation}
with KL-divergence regularisation:
\begin{equation}
    \mathcal{L}_{\text{KL}} = -\frac{\lambda_{\text{KL}}}{2} \sum_{j=1}^{d_m} \left(1 + \log\sigma_j^2 - \mu_j^2 - \sigma_j^2\right), \quad \lambda_{\text{KL}} = 10^{-3}
    \label{eq:kl}
\end{equation}

The VAE formulation provides three benefits: (i)~it structures the latent space for smooth interpolation between respiratory states, (ii)~it provides implicit denoising by preventing the encoder from encoding acquisition noise as motion information, and (iii)~the posterior variance offers a built-in uncertainty estimate for each motion code.

\subsection{Conditional deformation network}\label{sec:methods_deform}

The deformation network $D_\psi(\mathbf{x}, \mathbf{z}_m): \mathbb{R}^3 \times \mathbb{R}^{d_m} \to \mathbb{R}^3$ predicts a DVF mapping coordinates from the current respiratory state to the canonical frame:
\begin{equation}
    \mathbf{x}_{\text{ref}} = \mathbf{x} + D_\psi(\mathbf{x}, \mathbf{z}_m)
    \label{eq:warp}
\end{equation}

\subsubsection{Architecture}

A 4-layer SIREN~\citep{sitzmann2020siren} with 256 hidden units and $\omega_0 = 7$, with Tanhshrink final activation to bias the DVF toward sparse displacements (voxels not needing motion pushed toward zero).

\subsubsection{Motion code conditioning}

We investigate three conditioning strategies:

\textbf{Concatenation.} The motion code is broadcast and concatenated with spatial coordinates as input: $\text{input} = [\mathbf{x}; \mathbf{z}_m] \in \mathbb{R}^{3+d_m}$.

\textbf{Feature-wise Linear Modulation (FiLM)}~\citep{perez2018film}. The motion code generates per-layer scale and shift parameters:
\begin{equation}
    [\bm{\gamma}_l, \bm{\beta}_l] = \text{Linear}_l(\mathbf{z}_m), \quad h_l = \sin\!\left(\omega_0 \left(\bm{\gamma}_l \odot (W_l h_{l-1} + b_l) + \bm{\beta}_l\right)\right)
    \label{eq:film}
\end{equation}

\textbf{Adaptive multi-scale injection (default).} At layers 1--3, the motion code is projected through a dedicated linear layer followed by $\tanh$ activation, then added to the hidden representation:
\begin{equation}
    h_l = \sin(\omega_0 W_l h_{l-1}) + \tanh(\text{Linear}_l(\mathbf{z}_m))
\end{equation}
This additive injection at multiple depths allows coarse respiratory displacement to be captured by early layers while later layers refine local deformations, providing a balance between the expressiveness of FiLM and the stability of concatenation.

\subsubsection{DVF normalisation}

Network outputs are normalised to the $[-1,1]^3$ coordinate space: $\mathbf{d}_{\text{norm}} = 2\,\mathbf{d}_{\text{raw}} / \mathbf{S}$, where $\mathbf{S} = (H, W, D)$ is the volume shape in pixels and the division is element-wise. This normalisation ensures that the displacement magnitude is commensurate with the coordinate range regardless of anisotropic voxel dimensions.

\subsection{Bidirectional deformation with cycle consistency}\label{sec:methods_bidir}

We train both a forward DVF ($D_\psi^{\text{fwd}}$: frame $\to$ canonical) and an inverse DVF ($D_\psi^{\text{inv}}$: canonical $\to$ frame), sharing the same architecture but with independent parameters. Cycle-consistency loss enforces mutual invertibility:
\begin{equation}
    \mathcal{L}_{\text{cycle}} = \frac{\lambda_{\text{cyc}}}{N} \sum_{i} \left\| \mathbf{x}_i - \left(D^{\text{inv}}(\mathbf{x}_i^{\text{ref}}, \mathbf{z}_m) + \mathbf{x}_i^{\text{ref}}\right) \right\|_1, \quad \lambda_{\text{cyc}} = 0.1
    \label{eq:cycle}
\end{equation}
where $\mathbf{x}_i^{\text{ref}} = \mathbf{x}_i + D^{\text{fwd}}(\mathbf{x}_i, \mathbf{z}_m)$. The motivation for training a bidirectional DVF is downstream clinical utility (template-to-frame warping for dose accumulation and frame-to-frame motion via composition) rather than improved reconstruction fidelity; we defer the empirical comparison and the detailed downstream-utility argument to Sections~\ref{sec:results_bidir} and~\ref{sec:discussion}.

\subsection{Physics-aware thick-slice forward model}\label{sec:methods_thickslice}

\subsubsection{Motivation}

Clinical MRI slices integrate signal over a finite excitation profile (typically \SI{3}{\milli\metre}--\SI{7}{\milli\metre} slice thickness). Training an INR to match these partial-volume-averaged measurements at point coordinates creates a model mismatch: the INR is forced to represent blurred measurements as if they were point samples, degrading reconstruction fidelity. We address this by supervising the INR through a physics-aware forward model that accounts for the slice profile, improving the accuracy of the learned representation.

\subsubsection{Forward model}

The acquired intensity at position $\mathbf{x}$ with slice centre $z_c$ is modelled as:
\begin{equation}
    I_{\text{acq}}(\mathbf{x}) = \int_{z_c - \delta/2}^{z_c + \delta/2} P(z - z_c) \cdot f_\theta\!\left(\mathbf{x}_{z} + D_\psi(\mathbf{x}_{z}, \mathbf{z}_m)\right) dz
    \label{eq:thick_forward}
\end{equation}
where $\delta$ is the slice thickness, $P(\cdot)$ is the Gaussian excitation profile ($\sigma = \delta/4$), and $\mathbf{x}_{z}$ denotes the coordinate with depth set to $z$. Crucially, the DVF is evaluated independently at each depth position within the 
slice, correctly accounting for the fact that, \emph{at the same acquisition 
instant}, tissue at different through-plane depths can have different 
displacement vectors. For example, a diaphragm voxel at the anterior edge of
a 7\,mm coronal slice may have moved by a different cranio-caudal amount than a 
voxel at the posterior edge. This effect becomes non-negligible when slice 
thickness is comparable to the local spatial gradient of respiratory 
displacement.

\subsubsection{Gauss--Legendre quadrature}

The integral is approximated using 5-point Gauss--Legendre quadrature (nodes and weights in \ref{app:hyperparams}). The $S$-fold computational overhead occurs only during training; at inference, the INR is queried at original point coordinates.

\subsection{Loss functions}\label{sec:methods_loss}

The total training loss is:
\begin{equation}
    \mathcal{L} = \mathcal{L}_{\text{photo}} + w_s \cdot \mathcal{L}_{\text{reg}} + \lambda_{\text{KL}} \mathcal{L}_{\text{KL}} + \lambda_{\text{cyc}} \mathcal{L}_{\text{cycle}} + \lambda_{\text{inv}} \mathcal{L}_{\text{inv}}
    \label{eq:total_loss}
\end{equation}

Here $\mathcal{L}_{\text{reg}} = \lambda_{\text{jac}}\mathcal{L}_{\text{jac}} +
\lambda_{\text{mag}}\mathcal{L}_{\text{mag}}$ bundles the two DVF regularisers defined
below, and $\mathcal{L}_{\text{inv}}$ applies the same Jacobian penalty to the
inverse DVF (Eq.~\ref{eq:jac}) with weight $\lambda_{\text{inv}} = 10^{-4}$. The
scheduling factor $w_s$ is defined at the end of this section.

\textbf{Photometric loss.} L1 reconstruction using the thick-slice forward model:
\begin{equation}
    \mathcal{L}_{\text{photo}} = \frac{1}{N} \sum_{i=1}^{N} |I_{\text{thick}}(\mathbf{x}_i) - I_{\text{gt}}(\mathbf{x}_i)|
\end{equation}

\textbf{Jacobian regularisation.} Penalises deviation from volume preservation of the DVF:
\begin{equation}
    \mathcal{L}_{\text{jac}} = \frac{1}{N} \sum_{i} \left(\hat{d}_i + \frac{1}{\hat{d}_i} - 2\right), \quad \hat{d}_i = \text{clamp}(\det(\mathbf{J}_i),\, 0.1,\, 10), \quad \lambda_{\text{jac}} = 10^{-4}
    \label{eq:jac}
\end{equation}
where $\mathbf{J}_i = \mathbf{I} + \nabla D(\mathbf{x}_i)$ is the Jacobian of the deformation at point $i$. The determinant is clamped to $[0.1,\, 10]$ for numerical stability. This loss equals zero when $\det(\mathbf{J}) = 1$ (volume preservation) and increases for deviations in either direction, providing regularisation against both excessive compression and expansion. We note that this regulariser promotes near-volume-preserving deformations; for tissue that genuinely changes volume during breathing---most notably the lung parenchyma, whose air content varies substantially over the respiratory cycle---strict volume preservation does not hold, and a gradient-based smoothness regulariser may in principle be more appropriate. We deliberately apply only a small weight ($\lambda_{\text{jac}} = 10^{-4}$), so the term acts as a weak prior rather than a hard volume-preservation constraint and does not prevent the network from representing genuine lung volume change; it serves chiefly to suppress folding and implausible local distortions. The dominant respiratory excursion captured in our thoraco-abdominal data is the cranio-caudal motion of the diaphragm, liver, kidneys, and the chest and abdominal walls, for which near-incompressible soft-tissue deformation is a reasonable prior.

\textbf{DVF magnitude penalty.} L2 penalty on displacement magnitudes: $\mathcal{L}_{\text{mag}} = \|\mathbf{d}\|^2 / N$, $\lambda_{\text{mag}} = 0.01$, preventing unrealistically large deformations, particularly for rigid structures.

\textbf{KL divergence.} As defined in Eq.~\eqref{eq:kl}.

\textbf{Cycle consistency.} As defined in Eq.~\eqref{eq:cycle}. The inverse DVF is also regularised with the same Jacobian loss ($\lambda_{\text{inv}} = 10^{-4}$).

\subsection{Training and inference}\label{sec:methods_training}

\subsubsection{Joint optimisation}

All networks are optimised jointly with Adam~\citep{kingma2015adam} using module-specific learning rates (\ref{app:hyperparams}). Per training step, $B = 8$ slices are sampled and $N = 10{,}000$ coordinate points are randomly sub-sampled. Training details including the alternating evaluation-mode schedule are provided in \ref{app:hyperparams}.

\subsubsection{Inference}

At inference, a full 3D volume is reconstructed for each input 2D slice by: 
(i)~encoding the single acquired slice $S_n$ together with its slice-position index 
$k_n$ to obtain $\mathbf{z}_m = \bm{\mu}$ (posterior mean, no sampling at inference), 
(ii)~generating a dense 3D coordinate grid, (iii)~computing the DVF via $D_\psi$, 
and (iv)~querying $f_\theta$ at warped coordinates.


\section{Experiments}\label{sec:experiments}

\subsection{Datasets}\label{sec:exp_data}

\subsubsection{Dataset 1: \SI{1.5}{\tesla} clinical scanner}

The primary dataset comprises 5 healthy volunteers scanned on a \SI{1.5}{\tesla} MAGNETOM Aera scanner (Siemens Healthineers, Erlangen, Germany) using a prospective slice interleaving protocol~\citep{vonsiebenthal2007}. Each volunteer was scanned in three consecutive imaging sessions, with each session lasting around 10 minutes. Dynamic imaging employed a prospective slice-interleaved balanced steady-state free precession (bSSFP) protocol with Cartesian k-space sampling rather than repeated imaging at a single fixed slice. The protocol alternated between coronal image slices covering the whole body region and a sagittal navigator slice at a fixed position through the apex of the diaphragm. Each session comprised interleaved coronal 2D slices at up to $D = 64$ usable spatial positions ($D$ ranges from 36 to 64 across cases after preprocessing) covering the thoraco-abdominal region (chest and abdomen), acquired over 10--12 breathing cycles, yielding 768 slices per session. In-plane resolution: $\SIrange{1.04}{1.28}{\milli\metre} \times \SIrange{1.04}{1.28}{\milli\metre}$; slice thickness: \SIrange{3.64}{3.99}{\milli\metre}; flip angle: \SIrange{45}{53}{\degree}; echo time: \SI{1.09}{\milli\second}; acquisition time per slice: \SIrange{313}{331}{\milli\second}. Parallel imaging used GRAPPA factor 2 with 24 reference lines and partial Fourier 7/8.

We emphasise that, although these data were acquired with the
interleaved-navigator sequence of~\citet{vonsiebenthal2007}, we do
\emph{not} run the conventional navigator-based phase-sorting 4D
reconstruction at any stage, and we do not use any navigator-sorted volume
as training data or as a ground-truth reference. The model is trained
directly on the acquired coronal 2D slices, and each per-frame 3D volume is
estimated from a single acquired slice by the INR
(Section~\ref{sec:methods}). The interleaved sagittal navigator slices were
excluded from training and reserved solely as an independent reference for
evaluating the learned geometry and motion representation via
diaphragm-landmark tracking (Section~\ref{sec:exp_metrics}) and
motion-code surrogate correlation
(Section~\ref{sec:results_main}); the reference for landmark tracking is the
acquired navigator slice itself, not a reconstructed volume. All images were preprocessed by DICOM extraction, vertical flipping to radiological convention, cropping to $240 \times 290$ pixels, and window/level normalisation ($W = 300$, $L = 120$).

\textbf{Data partitioning.} Each session is split temporally: the first 9--11 breathing cycles for training; the final cycle (one slice per spatial position) for evaluation. This evaluates generalisation to unseen respiratory states. The three sessions per case enable assessment of inter-session reproducibility.

\subsubsection{Dataset 2: Low-field open-bore MRI}\label{sec:data_lowfield}

To demonstrate cross-field-strength and cross-position generalisability,
we evaluate on 10 acquisitions from five healthy volunteers scanned on a
\SI{0.5}{\tesla} MROpen system (ASG superconductors, Genova, Italy). This open-bore geometry
accommodates both supine (conventional) and upright (standing)
patient positioning. For each volunteer, we analyse one supine and one
upright scan, yielding $5 \times 2 = 10$ datasets. 

Data were acquired using a fast 2D HASTE protocol covering the thoracic-abdominal region ($D \in \{17, 19\}$ coronal slices, slice thickness \SI{12}{\milli\metre}, inter-slice spacing \SIrange{12}{13}{\milli\metre}, in-plane resolution $\sim 1.5$--$2.0$\,mm, temporal resolution $\approx \SI{0.5}{\second}$). The volume is acquired using an interleaved multi-slice pattern: rather than sequential scanning, readouts alternate between the lower and upper halves of the cranio-caudal range, starting from the central slice and progressing outward. This non-monotonic, zig-zag sweep is continuously repeated over a $\sim 10$~min recording, ensuring each spatial location is revisited multiple times across a full distribution of respiratory phases. Since no navigator or self-gating signal is available, our proposed image-driven encoder is designed to operate directly on this irregular temporal acquisition pattern. Finally, the substantially lower SNR of this \SI{0.5}{\tesla} data provides a challenging test case for both reconstruction quality and our implicit denoising hypothesis.

\textbf{Data partitioning.} Because the \SI{0.5}{\tesla} platform offers
no navigator-derived ground truth and no per-voxel volumetric reference,
each scan is split temporally: all but the last 50 acquired timepoints of
the multi-cycle sweep are used for training, and the final 50 timepoints
form the held-out evaluation window. Because the HASTE protocol revisits
each of the $D \in \{17, 19\}$ slice positions many times across the
$\sim 10$~min recording, this window contains roughly $50/D \approx
2.5$--$3$ revisits per spatial location, covering a full distribution of
respiratory phases unseen during training.
Image-quality metrics
are reported as slice-wise self-consistency between the INR-reconstructed
slice and the acquired 2D slice on the held-out portion
(Section~\ref{sec:exp_metrics}). For the implicit-denoising analysis
(Section~\ref{sec:results_lowfield_denoising}) we additionally use 50  acquired/reconstructed frame pairs per scan.

\subsection{Evaluation strategy}\label{sec:exp_baselines}

Rather than comparing against surrogate-dependent baselines that require navigator signals unavailable on many clinical platforms, we adopt systematic ablation as our primary benchmarking strategy (Section~\ref{sec:exp_ablation}). Each ablation variant isolates the contribution of a specific architectural component (template INR backbone, motion encoder design, deformation conditioning, bidirectional DVF, and physics-aware thick-slice supervision), providing insight into how each design choice affects reconstruction quality, geometric accuracy, and DVF regularity. The full proposed configuration serves as the reference against which all variants are compared.

In addition to component ablation, we evaluate the full pipeline along
four further axes that are reported in Section~\ref{sec:results}:
(i)~\emph{cross-platform self-consistency} on the \SI{0.5}{\tesla}
HASTE dataset, evaluated slice-wise on the held-out portion of the
multi-cycle sweep (Section~\ref{sec:results_lowfield});
(ii)~\emph{implicit denoising} via single-slice local SNR on matched
acquired/reconstructed frame pairs
(Section~\ref{sec:results_lowfield_denoising});
(iii)~\emph{DVF regularity} via Jacobian statistics on both datasets
and inverse-consistency error on the five bSSFP ablation runs
(Section~\ref{sec:results_dvf}); and
(iv)~\emph{computational performance}, measured as wall-clock
patient-specific training time and per-volume inference latency on a
single RTX 4090 GPU across all 15 bSSFP and 10 HASTE runs
(Section~\ref{sec:results_compute}).

\subsection{Evaluation metrics}\label{sec:exp_metrics}

\subsubsection{Image quality metrics}

Reconstruction quality is assessed on held-out evaluation slices using
peak signal-to-noise ratio (PSNR), structural similarity index
(SSIM)~\citep{wang2004ssim}, and mean absolute error (MAE). Both
datasets acquire only 2D slices, so no per-voxel volumetric reference
is available on either platform; all image-quality metrics are
therefore reported as \emph{slice-wise self-consistency} between the
INR-reconstructed slice and the acquired 2D slice. The two protocols
differ only in how the held-out window is selected, reflecting the
different temporal structure of the two acquisitions:

\begin{itemize}
    \item \textbf{\SI{1.5}{\tesla} bSSFP dataset.} The held-out window
    is the final breathing cycle of each session, which by
    construction visits every spatial position exactly once, so each
    case--session contributes up to $D$ paired comparisons (with
    $D \in [36, 64]$, see Section~\ref{sec:exp_data}).
    \item \textbf{\SI{0.5}{\tesla} HASTE dataset.} The held-out window
    is the last $50$ timepoints of the multi-cycle sweep. Because the
    HASTE protocol revisits each of the $D \in \{17, 19\}$ slice
    positions many times across the $\sim 10$~min recording, this
    window contains roughly $50/D \approx 2.5$--$3$ revisits per
    spatial location.
\end{itemize}

\subsubsection{Geometric accuracy metrics}

\textbf{Diaphragm-landmark tracking MAE.} (\emph{For the \SI{1.5}{\tesla} bSSFP dataset
only})
Five landmarks distributed across the diaphragm
dome are independently tracked on both the reconstructed and the reference (acquired)
sagittal navigator-plane slice images using CoTracker~\citep{karaev2024cotracker},
a transformer-based general-purpose video point tracker. The reconstructed
sagittal navigator plane is obtained by querying the INR at the same
spatial location as the acquired navigator slice; the reference is the
acquired navigator series itself. CoTracker was originally developed for natural RGB video. We apply it off-label to
grayscale 2D MR slice sequences after single-channel-to-RGB replication and
verified qualitatively that its predicted tracks remain visually consistent with
the diaphragm edge across breathing cycles on several test sequences. We
intentionally use \emph{the same} CoTracker configuration on both the reconstructed
and the reference sequence, so that any systematic tracking bias is cancelled in
the MAE difference; nonetheless, absolute landmark MAE values should be interpreted
as a relative comparison between methods rather than as a calibrated geometric
ground truth.

We use two distinct evaluation windows depending on the analysis:
\begin{itemize}
    \item \textbf{Full-sequence trajectory.} For each of the 15 bSSFP runs
    of the proposed full configuration we track all five diaphragm
    landmarks over the entire 768-frame reconstructed-vs-reference
    navigator series, enabling motion-profile analysis across all
    breathing cycles. This is the protocol used in
    Table~\ref{tab:per_case} (per-case breakdown).
    \item \textbf{Held-out cycle.} For the component ablations
    (Table~\ref{tab:ablation_summary}) the
    landmark MAE is restricted to the same held-out final breathing
    cycle used for the image-quality metrics, so that intensity and
    geometry metrics share a common temporal window.
\end{itemize}

In both cases, we report two summaries averaged over the five landmarks
and all evaluated frames: the cranio-caudal component $\text{MAE}_y$
(millimetres) and the 2-D Euclidean MAE in the image plane. Because
respiratory excursion is overwhelmingly cranio-caudal, $\text{MAE}_y$ is
the stricter of the two. We restrict landmark evaluation to the
diaphragm because it is the only thoraco-abdominal structure for which a
dense set of unambiguously identifiable, automatically-trackable points
exists across the full sequence; no Dice- or Hausdorff-style organ-mask
metrics are used.

\textbf{Per-case breathing-cycle window.} For the qualitative per-case
respiratory tracking of Section~\ref{sec:results_qualitative} (and
Fig.~\ref{fig:motion_cycle}), a single 40-frame breathing cycle is
selected automatically per case from the cranio-caudal landmark trace
$y(t)$: starting from the latest end-expiration peak, we extend backward
by $\sim 40$ frames to span one complete inspiration--expiration cycle.
The same 40-frame window is used for visualisation and for the
cycle-averaged $\text{MAE}_y$ (cranio-caudal) annotated in
Fig.~\ref{fig:motion_cycle}, so the figures and the cycle-averaged
metric refer to the same automatically selected frames.

\subsubsection{DVF quality metrics}

\textbf{Minimum Jacobian determinant.} We report
$\min_{\mathbf{x}} \det(\mathbf{J}(\mathbf{x}))$ of the estimated forward
DVF over the evaluation grid as the worst-case local
volume-preservation indicator; values close to 1 indicate
volume-preserving deformation, while values approaching or below 0 would
indicate local collapse or folding.

\textbf{Folding percentage.} Fraction of voxels with
$\det(\mathbf{J}) \leq 0$, quantifying topological violations.

\textbf{Inverse consistency error (ICE).}
$\| \mathbf{x} - (\mathbf{x} + D^{\text{fwd}}(\mathbf{x}) + D^{\text{inv}}(\mathbf{x} + D^{\text{fwd}}(\mathbf{x}))) \|$,
measuring consistency of bidirectional DVFs. ICE is reported only for
configurations that estimate both a forward and an inverse DVF;
unidirectional baselines are marked ``N/A''.

\textbf{Implicit-denoising SNR (\SI{0.5}{\tesla} only).} For the
low-field analysis of Section~\ref{sec:results_lowfield_denoising} we
additionally report a single-slice local SNR, defined as the mean
intensity in a manually placed liver ROI divided by the standard
deviation in a background ROI on the same slice. SNR is computed on the
matched acquired/reconstructed frame pairs ($n = 10$ scans, 50
frames per scan); we report both the per-frame mean and the per-scan
relative improvement (Reconstructed--Acquired)/Acquired.

\subsubsection{Statistical analysis}

Given the small sample size (5 subjects with 3 sessions each on the 1.5\,T 
dataset, 10 scans on the 0.5\,T dataset), we report mean $\pm$ standard deviation 
across runs and qualitative trends rather than formal significance tests; any 
comparison marked as "indistinguishable" or "within noise" in the results refers 
to overlapping standard deviations rather than a specific hypothesis test.

\subsection{Ablation study design}\label{sec:exp_ablation}
 
Systematic ablation studies evaluate the contribution of each architectural
choice. Unless otherwise specified, ablations are performed on the 5 cases
from the \SI{1.5}{\tesla} dataset (one session per case), with the full
model configuration as the reference.
 
\paragraph{Ablation 1: Template INR backbone}
We compare a 5-layer SIREN with residual connection at layer~4
($\omega_0 = 54$, 512 hidden units, $\sim$1.3M parameters) against the
default Instant-NGP backbone (16 hash levels with 2-feature-per-level
multi-resolution encoding feeding a 2-layer MLP, $\sim$1.1M parameters).
Both variants share identical motion encoders and deformation networks.
Evaluation criteria are final reconstruction quality (PSNR, SSIM, MAE),
held-out diaphragm-landmark MAE, and wall-clock patient-specific
training time for the full 10k-iteration schedule.
 
\paragraph{Ablation 2: Motion encoder architecture}
We isolate the variational bottleneck by comparing a deterministic
single-frame encoder (point estimate $\mathbf{z}_m = g_\phi(S, k)$) against
the proposed variational encoder (Gaussian posterior
$\mathcal{N}(\bm{\mu}, \bm{\sigma}^2)$ with KL regularisation). The two
share an identical convolutional backbone and slice-position conditioning
and differ only in the output head and the presence of the KL term.
Evaluation criteria are reconstruction quality (PSNR, SSIM),
held-out diaphragm-landmark MAE (2-D and cranio-caudal), and the
per-case Pearson correlation between the predicted cranio-caudal
landmark position and the navigator-derived reference, $r_\text{case}$,
which is reported here at the per-run level (15 runs) so that the
deterministic and variational encoders are compared on the same set of
runs. The motion-code surrogate correlation ($r_\text{case}$) reported in Table~\ref{tab:ablation_summary} (A2) characterises how well the default variational encoder captures the dominant respiratory mode.
 
\paragraph{Ablation 3: Deformation conditioning strategy}
Three strategies are compared: (i)~\emph{concatenation}, in which the
motion code is broadcast and concatenated with spatial coordinates as
input; (ii)~\emph{FiLM}, in which the motion code generates per-layer scale
and shift parameters that modulate the pre-activation signal; and
(iii)~\emph{adaptive multi-scale injection} (default), in which the motion
code is projected through a $\tanh$-gated linear layer at layers~1--3 and
added to the hidden representation. Evaluation criteria are reconstruction
quality (PSNR, SSIM), held-out diaphragm-landmark MAE, and DVF
regularity (inverse-consistency error and folding percentage).
 
\paragraph{Ablation 4: Bidirectional DVF and cycle consistency}
We compare a unidirectional DVF (forward only, frame~$\to$~canonical)
against the proposed bidirectional configuration with cycle-consistency
loss. Evaluation focuses on reconstruction quality and on DVF quality
metrics: minimum Jacobian determinant and folding percentage are
reported for both variants; inverse-consistency error is reported only
for the bidirectional variant, since the unidirectional baseline does
not estimate an inverse field.
 
\paragraph{Ablation 5: Physics-aware thick-slice supervision}
We compare na\"ive point-sampling supervision against the proposed
5-point Gauss--Legendre thick-slice supervision in which the DVF is
evaluated independently at each quadrature node. Evaluation focuses on
reconstruction quality assessed on held-out evaluation slices and on
reformatted coronal and sagittal views.

\subsection{Implementation details}\label{sec:exp_implementation}

All experiments are implemented in PyTorch and run on a single NVIDIA RTX 4090 GPU. The default hyperparameter configuration is summarised in Table~\ref{tab:hyperparams} (\ref{app:hyperparams}).


\section{Results}\label{sec:results}

\subsection{Component analysis on \SI{1.5}{\tesla} dataset}\label{sec:results_main}

We evaluate the contribution of each architectural component through systematic
ablation on the \SI{1.5}{\tesla} bSSFP dataset. The full proposed configuration is
the reference; each variant removes or replaces exactly one component. Given the
small sample size ($n=5$ cases, one session each), we report mean $\pm$ standard
deviation and qualitative trends rather than formal significance tests (see
Section~\ref{sec:exp_metrics}). Results for all five ablations are consolidated in
Table~\ref{tab:ablation_summary}; the subsections below interpret each.

\begin{table*}[htbp]
\centering
\caption{Compact ablation summary on the \SI{1.5}{\tesla} bSSFP dataset (five
cases, mean~$\pm$~std). Default variant in \textbf{bold}; best value per
group highlighted.}
\label{tab:ablation_summary}
\setlength{\tabcolsep}{2.8pt}
\scriptsize
\resizebox{\textwidth}{!}{%
\begin{tabular}{@{}llccccc@{}}
\toprule
Grp. & Variant & PSNR & SSIM & 2D MAE & $y$-MAE & Aux. \\
\midrule
\multirow{2}{*}{\textbf{A1} Backbone}
  & SirenRes                  & 24.96\,$\pm$\,0.66 & 0.864\,$\pm$\,0.014 & \textbf{3.95\,$\pm$\,1.00} & -- & 17.1\,min \\
  & \textbf{NGP}              & \textbf{25.42\,$\pm$\,0.72} & \textbf{0.872\,$\pm$\,0.015} & 4.13\,$\pm$\,1.11 & -- & \textbf{6.5\,min} \\
\midrule
\multirow{2}{*}{\textbf{A2} Encoder}
  & Deterministic             & \textbf{25.61\,$\pm$\,0.68} & 0.869\,$\pm$\,0.015 & 4.36\,$\pm$\,1.25 & 3.70\,$\pm$\,1.30 & $r=0.575$ \\
  & \textbf{VAE}              & 25.42\,$\pm$\,0.72 & \textbf{0.872\,$\pm$\,0.015} & \textbf{4.13\,$\pm$\,1.11} & \textbf{3.48\,$\pm$\,1.16} & $\mathbf{r=0.625}$ \\
\midrule
\multirow{3}{*}{\textbf{A3} Cond.}
  & Concatenation             & 22.93\,$\pm$\,0.45 & 0.824\,$\pm$\,0.016 & 5.13\,$\pm$\,1.39 & -- & ICE\,0.001 \\
  & FiLM                      & 22.37\,$\pm$\,0.69 & 0.812\,$\pm$\,0.026 & 5.40\,$\pm$\,1.76 & -- & ICE\,0.000 \\
  & \textbf{Adaptive}        & \textbf{25.42\,$\pm$\,0.72} & \textbf{0.872\,$\pm$\,0.015} & \textbf{4.13\,$\pm$\,1.11} & -- & ICE\,0.003 \\
\midrule
\multirow{2}{*}{\textbf{A4} DVF}
  & Unidirectional            & 25.39\,$\pm$\,0.66 & \textbf{0.872\,$\pm$\,0.016} & \textbf{4.06\,$\pm$\,1.06} & -- & $|J|_{\min}=0.983$ \\
  & \textbf{Bidirectional}   & \textbf{25.42\,$\pm$\,0.72} & \textbf{0.872\,$\pm$\,0.015} & 4.13\,$\pm$\,1.11 & -- & $\mathbf{|J|_{\min}=0.984}$ \\
\midrule
\multirow{2}{*}{\textbf{A5} Slice}
  & Point sampling            & \textbf{25.87\,$\pm$\,1.02} & \textbf{0.872\,$\pm$\,0.018} & \textbf{4.07\,$\pm$\,1.05} & -- & 4.2\,min \\
  & \textbf{Thick-slice GL}  & 25.42\,$\pm$\,0.72 & \textbf{0.872\,$\pm$\,0.015} & 4.13\,$\pm$\,1.11 & -- & \textbf{6.5\,min} \\
\bottomrule
\end{tabular}
}
\vspace{2pt}

{\raggedright\footnotesize
\textit{Aux.}: ablation-specific secondary metric. A1/A5: wall-clock training
time on one NVIDIA RTX 4090; A2: Pearson correlation $r$ between predicted
cranio-caudal diaphragm motion and the withheld navigator; A3:
inverse-consistency error (ICE, mm); A4: minimum Jacobian determinant
$|J|_{\min}$. All configurations achieve zero DVF folding.\par}
\end{table*}

\subsubsection{Template INR backbone}

Instant-NGP converges to a better intensity reconstruction in roughly one third of the training time of the sinusoidal SirenRes backbone. The sinusoidal activations of SirenRes do yield a marginally tighter landmark track (3.95 vs.\ 4.13\,mm), consistent with the smoother-frequency prior implicit to SIREN, but the 2.6$\times$ training speed-up of NGP is the dominant consideration for a patient-specific workflow that must be run per case. We therefore adopt NGP as the default template backbone.

\subsubsection{Motion encoder architecture}

The deterministic encoder achieves a marginally higher PSNR ($+0.19$\,dB), which we attribute to the absence of KL regularisation letting the latent capture reconstruction-helpful appearance cues at the cost of physiological relevance. The variational encoder instead yields a tighter diaphragm landmark track ($-0.23$\,mm on the 2-D MAE and $-0.22$\,mm cranio-caudally) and a stronger predicted-vs-reference diaphragm-$y$ correlation (0.625 vs.\ 0.575). This is the pattern one would expect if the VAE's KL bottleneck prevents the latent from encoding appearance noise as spurious motion while still describing the global respiratory state required by the deformation network, consistent with the VAE trading a fraction of pixel-level reconstruction accuracy for a more motion-faithful representation, although the differences are small relative to within-group standard deviations. Because a physiologically disentangled, smooth motion manifold is a stronger requirement for downstream clinical use than a small gain in intensity fidelity, the variational formulation is retained as the default.

\subsubsection{Deformation conditioning strategy}

The adaptive multi-scale injection strategy outperforms both alternatives by a substantial margin well beyond the 
one-standard-deviation band: $+2.5$\,dB PSNR over concatenation and $+3.1$\,dB over FiLM, with a concurrent $\approx$1\,mm reduction in diaphragm landmark error. The FiLM formulation, which modulates the pre-activation of every hidden layer uniformly, destabilises the sinusoidal activations and underperforms even the simple input-concatenation baseline; in contrast, adaptive injection at layers 1--3 with a $\tanh$-gated projection lets early layers capture the dominant respiratory mode while later layers continue to specialise anatomical detail. ICE remains sub-millimetric in all three variants, confirming that conditioning choice affects reconstruction fidelity without compromising bidirectional invertibility, and no variant produces any voxel with folding.

\subsubsection{Bidirectional DVF}\label{sec:results_bidir}

Image-quality and landmark metrics differ by less than one standard deviation between the two configurations, consistent with the fact that cycle-consistency is a regularisation constraint rather than a data term. The practical motivation for bidirectional DVF training is therefore not better reconstruction but downstream clinical utility. The inverse DVF enables template-to-frame warping needed for dose accumulation in MR-guided radiation therapy, and, because every inverse field is anchored to the same canonical template, it also turns the reconstruction into a complete inter-frame motion model: any frame-to-frame deformation $D^{i \to j}$ is recovered by composing $D^{\text{fwd}}(\cdot, \mathbf{z}_m^{(i)})$ with $D^{\text{inv}}(\cdot, \mathbf{z}_m^{(j)})$, with no additional training or pairwise registration. Both capabilities are produced essentially for free given that neither reconstruction fidelity nor deformation regularity is harmed. Both variants achieve effectively zero folding on the bSSFP dataset, so bidirectionality is retained as the default on the strength of the downstream-utility argument alone.

\subsubsection{Thick-slice supervision}

In-plane reconstruction metrics are marginally higher for point sampling, which is expected: point sampling is free to fit each observed slice exactly and therefore scores best on held-out in-plane PSNR even when the learned representation is not physically faithful. Thick-slice supervision, by contrast, forces the INR to be self-consistent with the physical slice excitation profile: the five-point Gauss--Legendre quadrature integrates the INR across the through-plane thickness during training and penalises the integrated, not the point-sampled, reconstruction. The quantitative cost is a modest $+2.3$\,min of training (a $1.5\times$ overhead from the five quadrature nodes) and a $\approx 0.45$\,dB drop in in-plane PSNR; the qualitative gain (Figure~\ref{fig:recon_examples}) is a clear reduction in coronal-reformat blurring through the slice-selection direction, where partial-volume averaging is concentrated. We therefore keep thick-slice supervision as the default because the target application is volumetric rather than slice-wise.

\subsection{Low-field open-bore MRI results}\label{sec:results_lowfield}

To test the portability claim made in Section~\ref{sec:intro}, we apply
the identical pipeline (same encoder, same INR, same deformation network,
same training recipe) to 10 \SI{0.5}{\tesla} HASTE acquisitions from a
different vendor (five volunteers, one supine and one upright scan each;
Section~\ref{sec:data_lowfield}). No hyperparameters, preprocessing steps,
or loss weights are changed from the \SI{1.5}{\tesla} defaults.

\begin{table}[htbp]
\centering
\caption{Reconstruction on the \SI{0.5}{\tesla} open-bore HASTE dataset
(five volunteers $\times$ supine/upright). PSNR / SSIM /
MAE are slice-wise self-consistency metrics (INR-reconstructed slice versus
acquired 2D slice on the held-out last 50 timepoints of the multi-cycle
sweep); $|J|_{\min}$ and folding\% quantify DVF regularity over the full
evaluation grid.}
\label{tab:results_lowfield}
\setlength{\tabcolsep}{3.2pt}
\footnotesize
\begin{tabular}{@{}lccccc@{}}
\toprule
Scan & PSNR $\uparrow$ & SSIM $\uparrow$ & MAE $\downarrow$ & $|J|_{\min}$ $\uparrow$ & Folding \% $\downarrow$ \\
\midrule
H1 Supine  & 27.70 & 0.706 & 0.032 & 0.974 & 0.000 \\
H1 Upright & 25.09 & 0.668 & 0.039 & 0.969 & 0.000 \\
H2 Supine  & 26.45 & 0.651 & 0.036 & 0.965 & 0.000 \\
H2 Upright & 26.59 & 0.650 & 0.034 & 0.965 & 0.000 \\
H3 Supine  & 27.08 & 0.657 & 0.034 & 0.960 & 0.000 \\
H3 Upright & 26.86 & 0.661 & 0.034 & 0.953 & 0.000 \\
H4 Supine  & 26.60 & 0.632 & 0.036 & 0.957 & 0.000 \\
H4 Upright & 26.97 & 0.652 & 0.034 & 0.945 & 0.000 \\
H5 Supine  & 26.55 & 0.653 & 0.036 & 0.957 & 0.000 \\
H5 Upright & 25.91 & 0.659 & 0.037 & 0.956 & 0.000 \\
\midrule
Mean $\pm$ Std & 26.58$\pm$0.67 & 0.659$\pm$0.018 & 0.035$\pm$0.002 & 0.960$\pm$0.008 & 0.000$\pm$0.000 \\
\bottomrule
\end{tabular}
\end{table}

\subsubsection{Supine position reconstruction}

On supine acquisitions (five scans) the framework achieves
$26.88 \pm 0.46$\,dB PSNR and $0.660 \pm 0.025$ SSIM at
self-consistency evaluation (Table~\ref{tab:results_lowfield}). The
supine PSNR is comparable to the \SI{1.5}{\tesla} bSSFP full-run mean
($26.58\pm2.04$\,dB; Table~\ref{tab:per_case}), a similarity that should
not be mistaken for equal image quality at the two field strengths: the
self-consistency metric rewards reconstruction fidelity against the acquired
input, and as Section~\ref{sec:results_lowfield_denoising} shows, the
reconstructed volumes at \SI{0.5}{\tesla} are substantially \emph{less}
noisy than the low-SNR input slices used for evaluation. DVF regularity is
preserved ($|J|_{\min} = 0.963 \pm 0.006$, zero folding voxels across all
five scans), confirming that the cycle-consistency regulariser transfers to
a different noise regime without retuning.

\subsubsection{Upright position reconstruction}

Upright acquisitions (five scans) produce $26.28 \pm 0.70$\,dB PSNR and
$0.658 \pm 0.006$ SSIM, overlapping the supine subset and qualitatively
similar in reconstructed volume sharpness. To our knowledge this is the
first demonstration of 4D MRI reconstruction from upright-posture data; it
is a direct consequence of the acquisition-agnostic design, since neither
external navigators nor k-space self-gating impose assumptions that couple
to body orientation. DVF regularity is again intact
($|J|_{\min} = 0.958 \pm 0.008$, no folding). The upright-supine comparison
within the same volunteer shows the expected posture-induced differences in
diaphragm resting position and in anterior--posterior organ shift, recovered
purely from the image data.

\begin{figure}[htbp]
    \centering
    \includegraphics[width=0.98\textwidth]{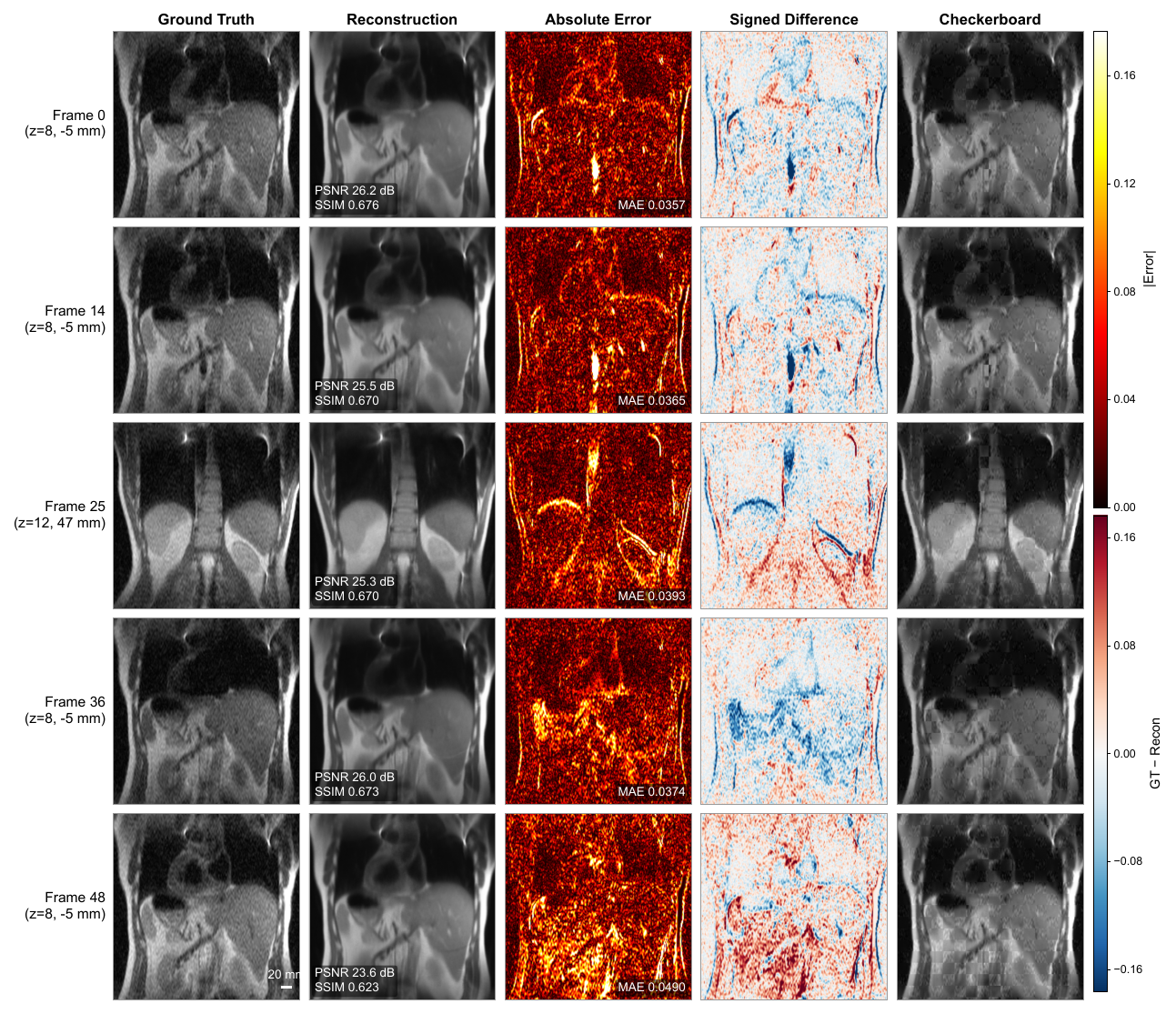}
    \caption{Slice-wise self-consistency on the \SI{0.5}{\tesla} HASTE
    dataset for a representative upright scan. Five frames from the held-out
    last-50-timepoint evaluation window are shown along rows; columns
    provide the acquired reference slice, INR reconstruction, absolute
    error, signed difference (Ref.$-$Recon) and a 16-pixel checkerboard
    overlay of reference and reconstruction. PSNR/SSIM/MAE are
    annotated per row.}
    \label{fig:lowfield_recon_vs_gt}
\end{figure}

Figure~\ref{fig:lowfield_recon_vs_gt} provides a visual companion to the
self-consistency numbers in Table~\ref{tab:results_lowfield} for one
upright scan: across the evaluation cycle the reconstructed slice tracks
the acquired ground-truth slice without visible registration artefacts,
and the checkerboard panel shows continuous anatomy across the
template/reconstruction blocks despite the low SNR.

\subsubsection{Implicit denoising effect}\label{sec:results_lowfield_denoising}

\begin{figure}[htbp]
    \centering
    \includegraphics[width=0.98\textwidth]{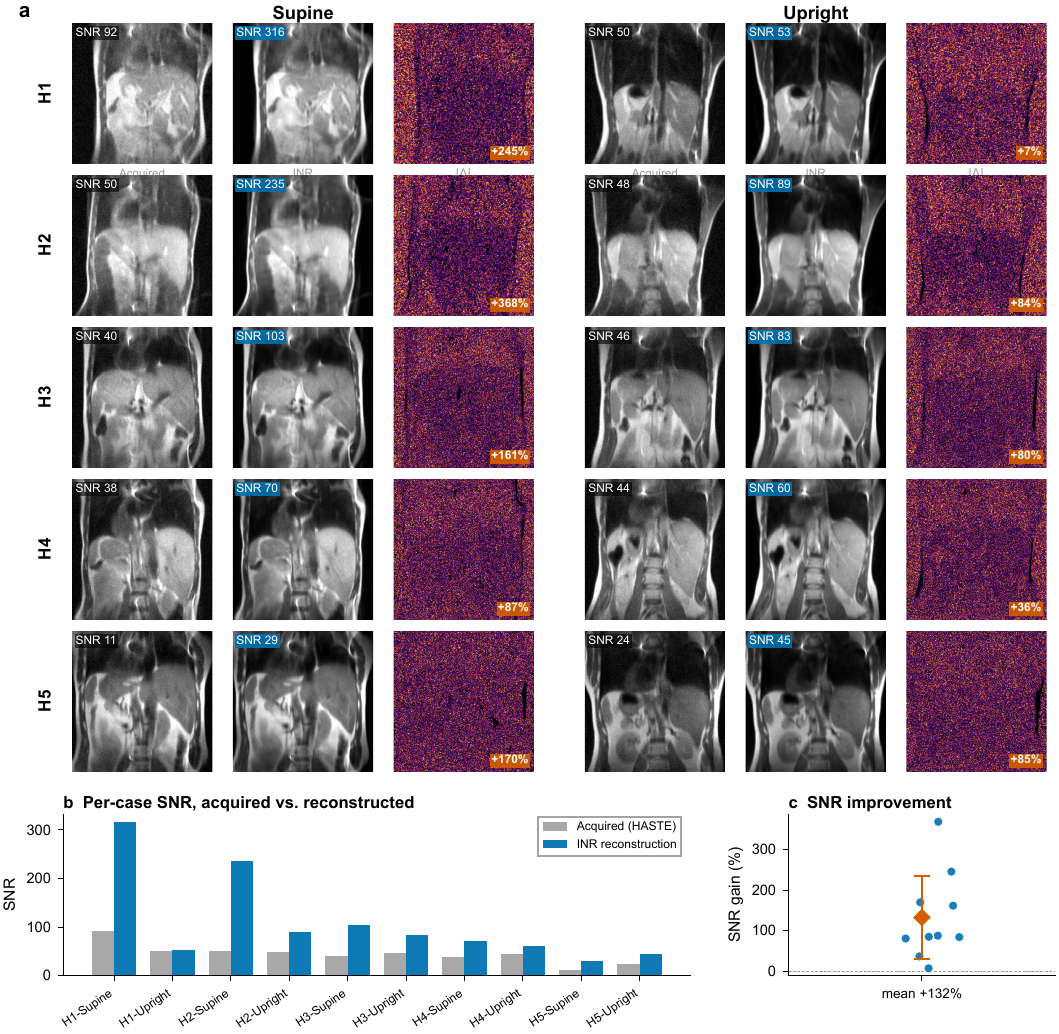}
    \caption{Implicit denoising on the \SI{0.5}{\tesla} HASTE dataset
    (10 scans). Rows correspond to anonymised volunteers H1--H5 and
    columns to supine/upright positioning; each case shows the acquired
    slice, the corresponding INR reconstruction and their absolute
    difference with local-background SNR badges. The scatter--bar panel at
    the bottom aggregates per-scan values across all cases.}
    \label{fig:denoising}
\end{figure}

A notable finding on the low-field data is that the INR template provides implicit denoising. Because each spatial position is revisited multiple times over repeated sweeps spanning many breathing cycles, the canonical template network $f_\theta$ is trained on redundant observations of the same anatomy corrupted by independent noise realisations. The INR naturally learns the common signal (anatomy) while averaging out the uncorrelated noise, producing a canonical volume with substantially higher SNR than any individual acquired slice. The deformation network then maps this clean template to each per-frame state, preserving the denoising benefit in the final reconstructed volumes.

This implicit denoising arises without any explicit denoising loss or noise
model; it is a natural consequence of the INR's smoothness bias and the
redundancy in multi-cycle acquisitions. Across all 10 scans, we measure the
single-slice local SNR (signal mean in a liver ROI divided by noise standard
deviation in a background ROI) on both the acquired 2D slice and the
corresponding INR-reconstructed slice, matched frame-by-frame. Averaged over
case means, SNR rises from $44.3 \pm 19.8$ on the acquired slices to
$108.6 \pm 88.0$ on the reconstructed slices. The per-scan relative
improvement is $+132.4 \pm 102.5\%$ ($n=10$ scans, $50$ frames per scan;
range $+7\%$ to $+368\%$, see Fig.~\ref{fig:denoising}). The variance
across scans reflects both baseline noise level and subject-specific anatomy:
the largest relative gain occurs in H2 supine ($+368\%$), whereas the
already-clean H1 upright scan shows only a small relative gain ($+7\%$).

\subsection{DVF quality analysis}\label{sec:results_dvf}

\begin{figure*}[htbp]
    \centering
    \includegraphics[width=\textwidth]{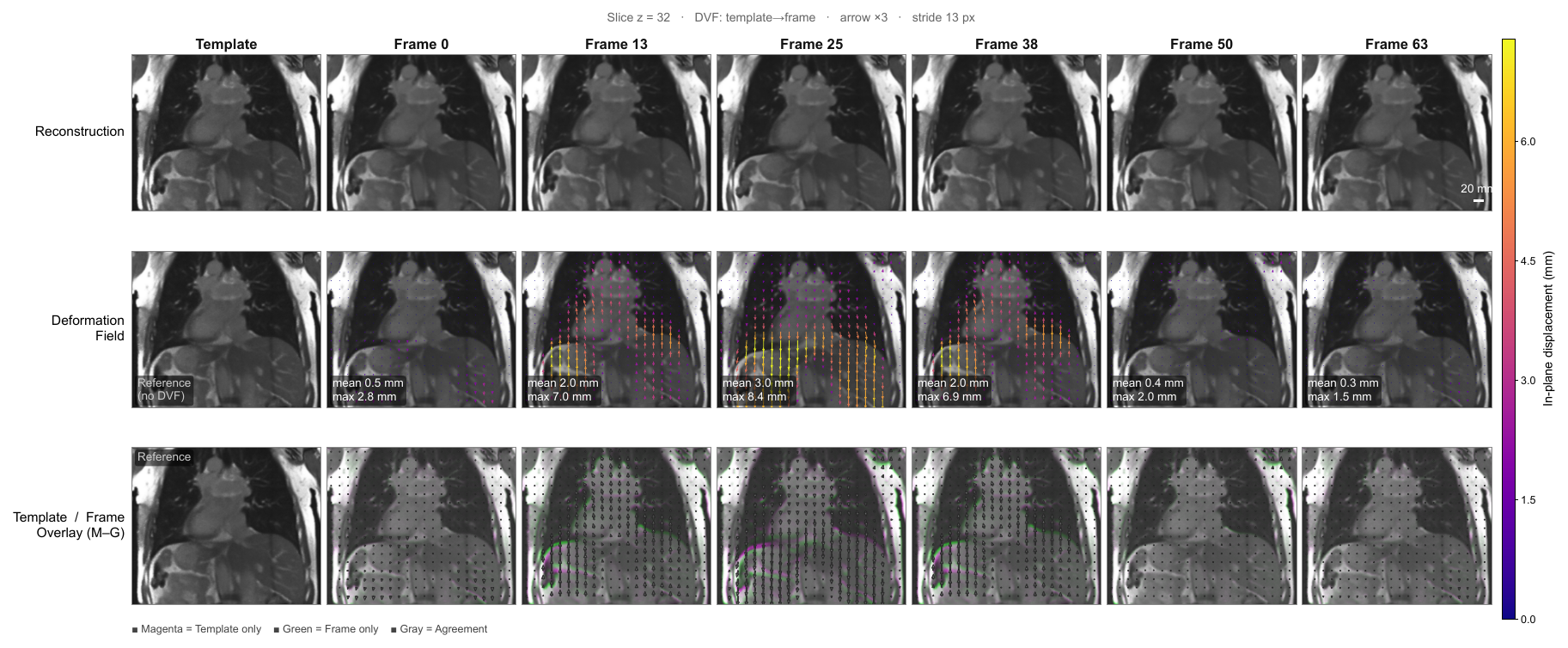}
    \caption{Temporal evolution of the predicted template-to-frame DVF
    for case C04, session~1, mid-coronal slice ($z=32$/64). Top row:
    reconstructed coronal frame at the canonical template and six evenly
    spaced respiratory phases. Middle row: the same frames overlaid with
    in-plane DVF arrows; arrow colour encodes in-plane displacement
    magnitude (mm) on the shared colour bar (right). Bottom row: a
    magenta--green alignment overlay between the canonical template and
    each warped frame; pure grey indicates well-aligned regions,
    magenta or green casts indicate residual mismatch.}
    \label{fig:dvf_temporal}
\end{figure*}

Quantitatively, the estimated DVFs are regular on both datasets: the minimum Jacobian determinant is $|J|_{\min} = 0.984 \pm 0.004$ on the \SI{1.5}{\tesla} bSSFP evaluation cycles and $|J|_{\min} = 0.960 \pm 0.008$ on the \SI{0.5}{\tesla} HASTE scans (Table~\ref{tab:ablation_summary},~\ref{tab:results_lowfield}). These Jacobian statistics are evaluated over the entire reconstruction grid spanning the full thoraco-abdominal field of view (including the lung), not over a single-organ region of interest; the near-unity minimum reflects that, with the deliberately small Jacobian weight (Section~\ref{sec:methods_loss}), the field stays close to volume-preserving in soft tissue while remaining free to represent lung volume change without folding. No folding voxels ($\det(\mathbf{J}) \le 0$) are observed in any of the scans for which Jacobian statistics are tabulated (5 bSSFP ablation runs in Table~\ref{tab:ablation_summary} and 10 HASTE scans in Table~\ref{tab:results_lowfield}); we
therefore observe no topological violations at the voxel grid, although we note
that this is a necessary but not sufficient condition for global diffeomorphism
of the continuous DVF. Inverse consistency, defined as the mean residual $\|\varphi \circ \varphi^{-1} - \mathrm{id}\|$ between the forward and inverse DVFs, is $\mathrm{ICE} = 0.003 \pm 0.002$\,mm with the default adaptive conditioning,
more than two orders of magnitude below the $\sim$1\,mm in-plane voxel pitch, confirming that the cycle-consistency loss produces genuinely bidirectional fields rather than two independent networks. Displacement is concentrated in the expected regions: large cranio-caudal excursions at the diaphragm and liver dome ($>15$\,mm), moderate excursions at the kidneys, and near-zero displacement at the spine and posterior body wall. Fig.~\ref{fig:dvf_temporal} shows that the predicted DVF varies smoothly across the held-out respiratory cycle: arrows reverse direction monotonically between end-expiration and end-inspiration, the template-to-frame alignment overlay (bottom row) collapses to neutral grey in the well-registered regions, and no abrupt jumps or folding patterns appear at any phase, consistent with the per-frame Jacobian statistics above.

\subsection{Qualitative results}\label{sec:results_qualitative}

\subsubsection{Per-case reconstruction quality}

\begin{figure*}[htbp]
    \centering
    \includegraphics[width=\textwidth]{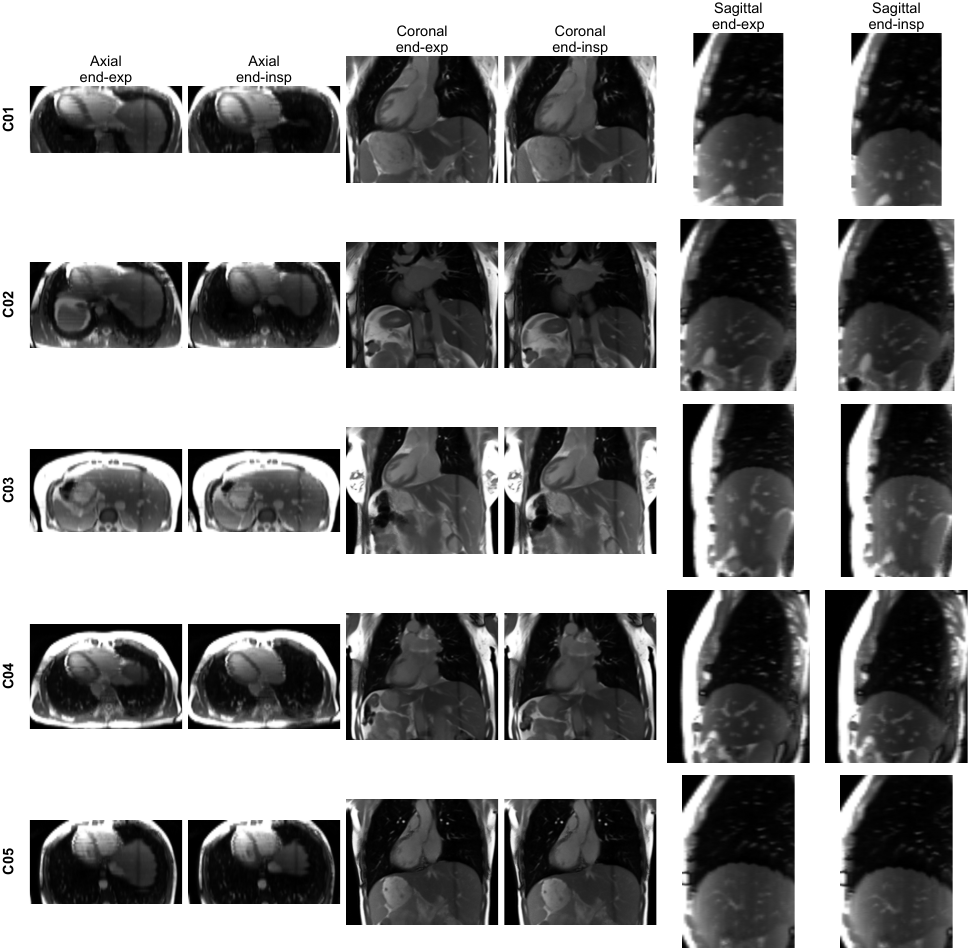}
    \caption{Per-case reconstruction quality on the \SI{1.5}{\tesla} bSSFP
    dataset (one volunteer per row, session~1). Columns show axial, coronal, and sagittal views at end-expiration (EE) and end-inspiration (EI), plus a per-pixel temporal MIP showing the diaphragm motion envelope.}
    \label{fig:recon_per_case}
\end{figure*}

Figure~\ref{fig:recon_per_case} shows the per-case reconstruction quality
across the five bSSFP volunteers in session~1. The framework produces
temporally coherent, sharply resolved 3-D volumes in every case despite
sizeable inter-subject differences in baseline field of view (H$\times$W
$=240\times290$; $D=36$--$64$), slice thickness (3.6--4.0\,mm), and
respiratory amplitude (see also Fig.~\ref{fig:motion_cycle}). The
end-expiration to end-inspiration contrast is visible in all three
views: the diaphragm dome is clearly more superior at EE than at EI in
the sagittal and coronal panels, and the same motion produces an
organ-boundary shift in the axial reformat that is cleanly resolved by
the predicted DVF. The cycle MIP panel is a compact qualitative summary
of motion range: a thin static band indicates low motion (e.g.\ C03), a
wide blurred band indicates deep breathing (C04, C05). Across all cases
the spine and posterior body wall are sharp throughout the MIP, which is
consistent with the near-zero displacement expected there and with the
Jacobian-regularity statistics of Section~\ref{sec:results_dvf}.

\subsubsection{Coronal MIPs of the reconstructed 3-D volumes}

\begin{figure*}[htbp]
    \centering
    \includegraphics[width=0.95\textwidth]{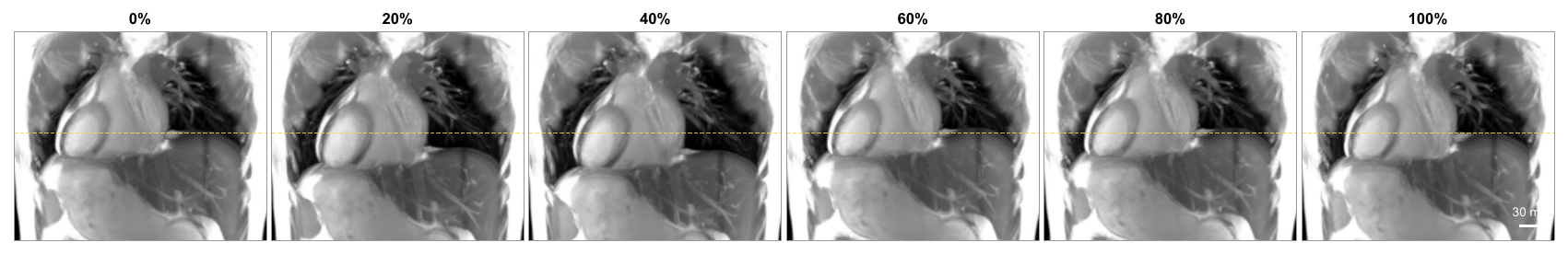}
    \caption{Coronal maximum-intensity projections (MIPs) of the
    reconstructed 3-D volume at five evenly spaced phases of the
    held-out respiratory cycle for case C04, session~1
    (\SI{1.5}{\tesla} bSSFP). Each panel projects an anterior--posterior
    sub-volume through the upper abdomen along the depth axis. Volume
    spacing $\approx 3.99\!\times\!1.28\!\times\!1.28$\,mm; the
    rightmost panel carries a \SI{30}{\milli\metre} scale bar. All
    panels share contrast settings (1\textsuperscript{st}--99.5\textsuperscript{th}
    percentile window, $\gamma = 0.65$).}
    \label{fig:mip_phases}
\end{figure*}

Figure~\ref{fig:mip_phases} shows coronal MIPs of the reconstruction at five respiratory phases for case C04, session~1. The diaphragm contour is a single sharp arc in every panel, organ boundaries are continuous, and phase-to-phase changes trace the expected cranio-caudal respiratory motion ($\sim$15--20\,mm), directly demonstrating volumetric conformity of the reconstructed 4D MRI.

\subsubsection{Methods comparison at matched respiratory states}

\begin{figure*}[htbp]
    \centering
    \includegraphics[width=0.98\textwidth]{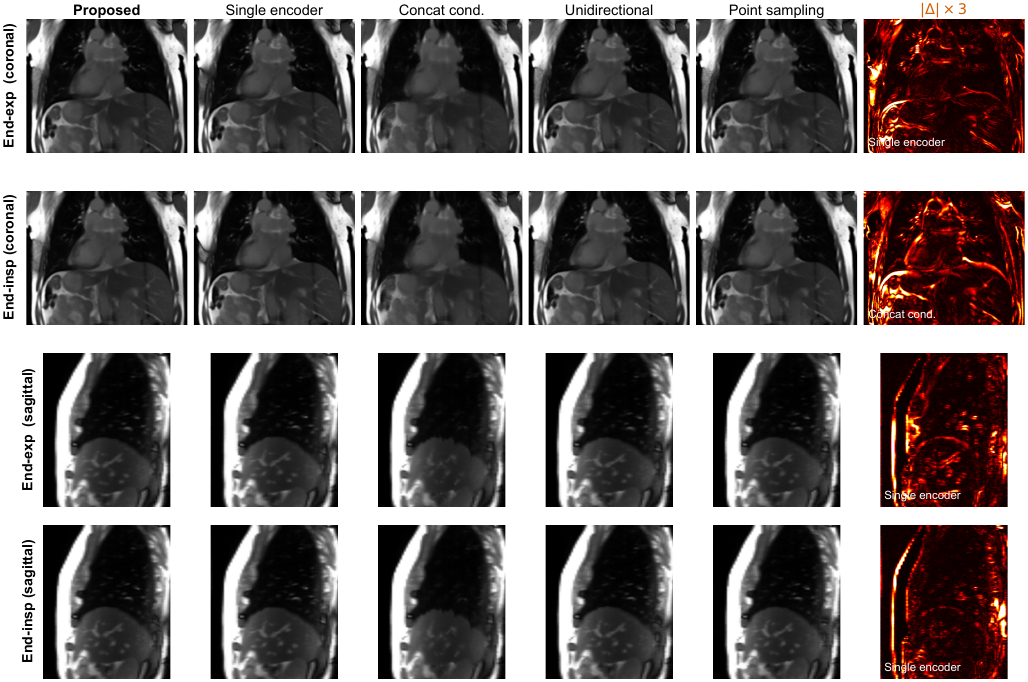}
    \caption{Ablation variants compared at matched respiratory states
    (case C04, session~1). Rows: end-expiration and end-inspiration in
    coronal and sagittal views (resampled to isotropic spacing).
    Columns: the proposed full configuration, selected ablation variants
    (deterministic encoder, concatenation conditioning, unidirectional
    DVF, point sampling), and a $3\times$-magnified absolute difference
    map between the proposed and the worst ablation at each row.}
    \label{fig:recon_examples}
\end{figure*}

Figure~\ref{fig:recon_examples} compares the full configuration against four ablation variants at two matched respiratory states (C04, session~1). The full method recovers the sharpest organ boundaries, particularly at the diaphragm and liver dome; ablations show blurring or slice-thickness-scale ringing. Residual error concentrates at the diaphragm--liver interface, consistent with quantitative results (Table~\ref{tab:ablation_summary}).

\subsubsection{Per-case respiratory cycle tracking}

\begin{figure*}[htbp]
    \centering
    \includegraphics[width=\textwidth]{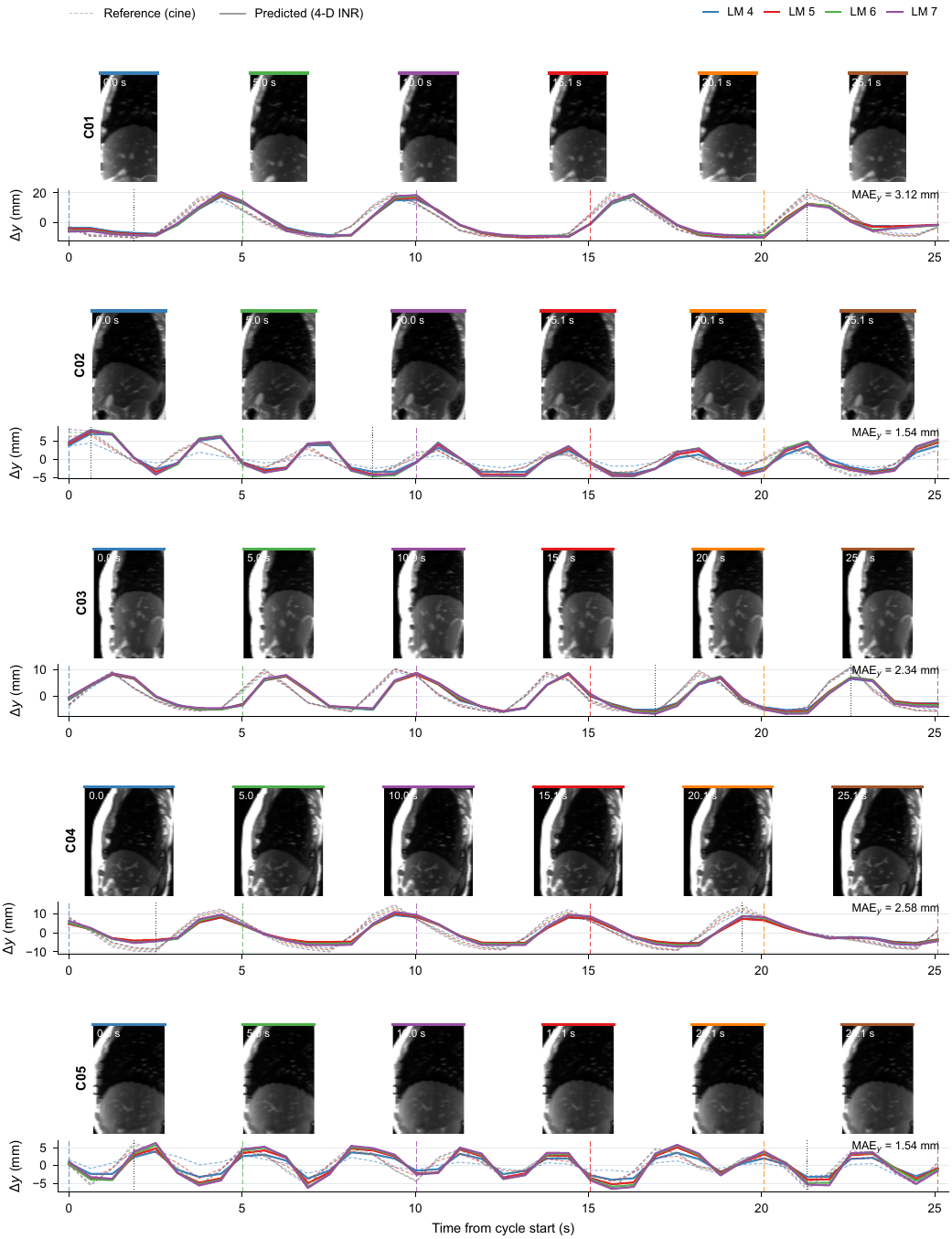}
    \caption{Per-case breathing-cycle tracking on the \SI{1.5}{\tesla}
    bSSFP dataset (five volunteers, session~1). For each case the top
    strip shows six sagittal reformats at evenly spaced respiratory phases;
    the bottom panel plots diaphragm landmark trajectories (thin dashed = CoTracker reference, thick solid = predicted).}
    \label{fig:motion_cycle}
\end{figure*}

Figure~\ref{fig:motion_cycle} shows per-case breathing cycles (C01--C05, session~1) to compare reconstructed and measured respiratory dynamics. The six sagittal reformats per case confirm temporal coherence: organ boundaries move smoothly and the overlaid landmark trajectories show close agreement between predicted and reference in both amplitude and phase.

\subsection{Computational performance}\label{sec:results_compute}

\begin{table}[htbp]
\centering
\caption{Computational performance of the proposed framework, measured on a single NVIDIA RTX 4090 GPU over the 15 \SI{1.5}{\tesla} bSSFP runs and 10 \SI{0.5}{\tesla} HASTE runs.}
\label{tab:compute}
\footnotesize
\begin{tabular}{lc}
\toprule
Stage & Time \\
\midrule
Preprocessing                       & None (operates on reconstructed 2D slices) \\
Training (per patient, 10k iter.)   & $6.2 \pm 0.1$\,min \\
Inference (per 3-D volume)          & $412 \pm 26$\,ms (0.5\,T HASTE) / $1187 \pm 1$\,ms (1.5\,T bSSFP) \\
\bottomrule
\end{tabular}
\end{table}

Training required $6.2 \pm 0.1$ minutes per patient for the default $10{,}000$-iteration schedule (mean across all 15 bSSFP and 10 HASTE runs; the ablation-batch timing of 6.5\,min in Table~\ref{tab:ablation_summary} (A1/A5) reflects a separate five-run session-1 batch on the same hardware). Reconstructing a full 3-D volume for any respiratory state required $412 \pm 26$\,ms on the \SI{0.5}{\tesla} HASTE acquisitions (volume dimensions $\sim 192 \times 192 \times 19$) and $1187 \pm 1$\,ms on the \SI{1.5}{\tesla} bSSFP acquisitions (volume dimensions $\sim 240 \times 290 \times 64$, i.e.\ about $6.4\times$ more voxels). Per-volume latency therefore increases with volume size, rather than being an architectural bottleneck. Compared with the respiratory timescale ($\sim 3$--$5$\,s per cycle, with gating decisions needed every few hundred milliseconds), the HASTE latency is well within the real-time regime; the bSSFP latency approaches it, and can be further reduced by re-training on a smaller volume grid or by using optimised batched INR rendering. We note that ``real time'' refers to inference speed, not the complete workflow: the per-patient training phase must precede online inference. In an MR-Linac context, this training could be performed during patient setup (typically 5--15 minutes).


\section{Discussion}\label{sec:discussion}

\subsection{What SIMPLE-4D changes}\label{sec:disc_portable}

The encoder ablation (Table~\ref{tab:ablation_summary}, A2) provides the
empirical check that this is the right trade-off. On the five-case ablation
subset, the variational encoder achieves a diaphragm-landmark correlation of
$r=0.625$ with the withheld navigator, versus $r=0.575$ for the deterministic
variant, confirming that nothing essential is lost by removing the surrogate.
Across all 15 \SI{1.5}{\tesla} bSSFP runs (5 volunteers $\times$ 3 sessions) the default VAE encoder achieves
$r_\text{case} = 0.68 \pm 0.14$, consistent with the ablation finding and
generalising it across subjects and sessions. The
32-dimensional code additionally carries information that a scalar
surrogate cannot represent, including hysteresis, differential motion, and
non-periodic dynamics~\citep{seppenwoolde2002,ge2013,dhont2018}.
The bidirectional DVF ablation (Table~\ref{tab:ablation_summary}, A4)
shows that in-plane image-quality and landmark metrics are within one
standard deviation between unidirectional and bidirectional
configurations; bidirectionality is therefore retained on the strength of
its downstream clinical utility (enabling dose accumulation and
inter-frame motion model composition via $D^{i \to j} = D^{\text{fwd}}_i
\circ D^{\text{inv}}_j$) rather than on the basis of superior
reconstruction metrics.

\subsection{The limits of acquisition agnosticism}\label{sec:disc_acqagnostic}

We stress that acquisition agnosticism is not the same as acquisition
invariance. SIMPLE-4D decouples from axes~1--6 of
Section~\ref{sec:intro} but inherits a residual dependence on image
contrast (bSSFP and HASTE are validated; extensions to T1-weighted
gradient-echo or contrast-enhanced protocols require dedicated evaluation),
receive-coil configuration, and the absolute SNR level set by field strength
and voxel size. The cross-platform validation quantifies this empirically:
reconstruction fidelity, deformation regularity, and the magnitude of the
implicit denoising benefit all vary between \SI{1.5}{\tesla} bSSFP and
\SI{0.5}{\tesla} HASTE in ways reported in
Section~\ref{sec:results_lowfield}. What is invariant across the two
platforms is the \emph{pipeline itself}: no code path changes between the
datasets.

\subsection{Low-field, upright, and where portability matters most}\label{sec:disc_lowfield}

Portability is most valuable in settings where conventional 4D MRI is hard
to deploy in practice. We stress that this is a question of practical
availability, not of physical possibility: the classical navigator-based
4D-MRI approach~\citep{vonsiebenthal2007} is itself essentially
field-strength agnostic and could in principle be ported to low-field
systems. What it requires, however, is an interleaved-navigator pulse
sequence---a vendor-specific sequence modification (axis~2)---together with
a maintained 1D respiratory surrogate (axis~4). On many low-field,
open-bore, and upright commercial platforms, including the \SI{0.5}{\tesla}
open-bore system used here, such a navigator sequence is not provided and
cannot be added without sequence-programming access, and upright
positioning additionally lacks compatible external respiratory hardware.
By operating purely in image space on the reconstructed 2D slices that
every scanner already produces, our approach sidesteps these practical
prerequisites rather than any fundamental limit, enabling volumetric 4D
reconstruction where the conventional acquisition infrastructure is
unavailable.

The implicit denoising effect on low-field data
(Fig.~\ref{fig:denoising}) is a secondary benefit of this design. Because the
canonical template is trained on many observations of the same anatomy with
independent noise realisations, and because the multi-resolution
hash-encoded INR is biased towards the lower-frequency components of the
target function before the higher-frequency
ones~\citep{rahaman2019spectral,tancik2020fourier}, the network learns the
underlying coherent anatomy well before it begins to fit the voxel-level
noise. Analogous implicit-denoising behaviour has been reported for
INR-based sparse-view MRI reconstruction~\citep{shen2022nerp} and 4D flow
MRI super-resolution~\citep{saitta2023inr4dflow}.

\subsection{Physics-aware thick-slice training}\label{sec:disc_thickslice}

The thick-slice forward model reduces through-plane blurring artefacts at organ boundaries (Figure~\ref{fig:recon_examples}); in-plane quantitative metrics favour point sampling (Table~\ref{tab:ablation_summary}, A5), suggesting the benefit is primarily qualitative in reformatted views.

\subsection{Implications for MR-guided radiotherapy}\label{sec:disc_mrrt}

SIMPLE-4D has three implications for MR-guided radiotherapy. First, it deploys on ViewRay MRIdian, Elekta Unity, or any MR-Linac platform without sequence modifications. Second, inference latency is within the respiratory timescale (Table~\ref{tab:compute}), enabling full-volume monitoring for beam-on gating decisions. Third, per-frame output enables breath-by-breath dose accumulation capturing inter-cycle variability. Upright reconstruction extends applicability to non-supine configurations.

\subsection{Limitations and future work}\label{sec:disc_limits}

\textbf{Image-domain operation.} Operating on reconstructed 2D slices
rather than raw k-space data means parallel-imaging artefacts from initial
slice reconstruction (e.g.\ GRAPPA) propagate into the 4D reconstruction.
An end-to-end k-space-to-volume approach~\citep{hammernik2023physics} could
improve fidelity but would sacrifice acquisition agnosticism.

\textbf{Per-patient training.} The patient-specific training paradigm
requires approximately 6\,minutes per subject, which must precede online
inference. For MR-Linac applications, this can be performed during patient
setup; for truly real-time deployment, transfer learning or meta-learning
across patients is a promising direction.

\textbf{Volumetric ground truth.} Both datasets acquire only 2D slices,
so no per-voxel volumetric reference is available on either platform.
Reconstruction fidelity is therefore assessed through slice-wise
self-consistency on held-out frames and through DVF regularity rather
than through direct volumetric comparison. A 3D breath-hold reference
provides only a static template and does not resolve the underlying
ill-posedness of dynamic volumetric validation. Two more informative
strategies are left to future work: (i)~scanning an MR-compatible
physical 4D phantom driven by a reproducible motion programme, which
would yield a true 3D$+$t ground truth against which both the
reconstructed volumes and the estimated DVFs can be compared, and
(ii)~simulated multi-slice acquisitions from an existing high-resolution
4D dataset (e.g.\ a 4D-CT or a densely sampled 4D-MRI), which would
enable controlled benchmarking against a known ground truth across
acquisition protocols, slice geometries, and SNR levels.

\textbf{Healthy volunteer validation.} While we validate across two field
strengths and positions, the target population for radiotherapy,
oncology patients, may exhibit irregular breathing patterns and
anatomical distortions from tumours. Patient-specific validation is the
natural next step.

\textbf{Single-valued deformation.} Our DVFs assume smooth, continuous
motion and do not explicitly model sliding at organ boundaries (e.g.\
lung--chest wall interface)~\citep{wu2008sliding}. For applications
requiring precise interfacial sliding, region-specific deformation models
may be needed.

\textbf{Cardiac and other motion types.} The motion-code framework is
not intrinsically limited to respiratory motion: the latent encoder can
in principle capture any periodic or quasi-periodic deformation present
in the image data, provided the training distribution covers it.
The current validation focuses on respiratory motion because the temporal resolution of these experimental datasets, which were acquired in the thoracic-abdominal region during
free-breathing, are dominated by this type of motion; other motion patterns
have not yet been validated.
Future work should extend the evaluation to datasets that include cardiac
motion, peristalsis, or voluntary body motion to confirm the generality
of the approach.
Cardiac motion, although smaller in amplitude, contributes to the organ
displacement in the thoracic-abdominal region; a multi-scale temporal
encoding could extend the motion manifold to capture both respiratory and
cardiac dynamics simultaneously, as explored
in~\citep{christodoulou2018multitasking,feng2018fiveDwholeheart}.
DREME-MR~\citep{shao2025dremeMR} demonstrates one such direction using
INR-based joint cardio-respiratory motion estimation with a
frequency-guided progressive training strategy.


\section{Conclusion}\label{sec:conclusion}
 
We have presented \textbf{SIMPLE-4D} (\textbf{S}urrogate-free,
\textbf{IM}plicit, \textbf{P}ortab\textbf{LE} 4D MRI), a software-first,
portable workflow for 4D MRI that reconstructs per-frame 3D volumes from
standard fast multi-slice 2D MRI, without non-Cartesian trajectories,
without pulse-sequence modifications, without external respiratory
hardware, and without any 1D respiratory surrogate. SIMPLE-4D combines an
acquisition-agnostic front end that consumes reconstructed 2D slices, an
image-driven variational single-frame motion encoder that extracts a
compact motion code from each 2D slice and its slice-position embedding,
and a continuous spatio-temporal INR reconstruction with a physics-aware
thick-slice forward model and bidirectional cycle-consistent DVFs.
Cross-platform validation at \SI{1.5}{\tesla} bSSFP and \SI{0.5}{\tesla}
open-bore HASTE in both supine and upright positions, with a single
unmodified SIMPLE-4D pipeline, confirms that the workflow is genuinely
platform-agnostic, and includes, to our knowledge, the first demonstration
of per-frame 4D reconstructed-volume respiratory MRI on a weight-bearing
upright open-bore low-field scanner from reconstructed 2D Cartesian slices
alone. The INR template additionally provides implicit
denoising on low-field data as a free by-product of multi-cycle redundancy.
By moving 4D MRI out of the acquisition and into software, SIMPLE-4D lowers
the deployment threshold for volumetric motion-resolved imaging on
existing clinical scanners, low-field MR-Linac platforms, open-bore and
upright systems, and emerging portable MRI deployments in which
conventional 4D MRI pipelines cannot run.


\section*{Funding}

This research was supported by the Swiss National Science Foundation (SNSF) under grants 212855 (EPIC-4DAPT) and 163330, and by the ETH Domain iDoc grant 2021-360 through Personalized Health and Related Technologies (PHRT). The funders had no role in the study design; the collection, analysis, and interpretation of data; the writing of the manuscript; or the decision to submit the article for publication.

\section*{Acknowledgments}

The authors sincerely thank Prof. Dr. Damien C. Weber for his clinical insight and management of the ethical studies. We are grateful to the PSI-CPT RTT team, especially Michael Zorneth, Stephanie Trösch, and Tiinaleena Lazic, for their assistance with data acquisition. Additionally, we thank Niels Oesingmann, Leonardo Bertora, Luisa Raimondo, and Anna Sampanai (ASG Superconductors) for their support with the \SI{0.5}{\tesla} HASTE acquisitions, and Oliver Bieri (University of Basel) for his contributions to the \SI{1.5}{\tesla} 4D MRI sequence development.

\section*{Data availability}

The MRI datasets analysed in this study cannot be made publicly available because they consist of identifiable human volunteer imaging data acquired under ethics approvals (EKNZ 2019-01060 and EKNZ 2023-00674) and the associated informed-consent agreements, which do not permit open distribution, and because of data-protection and privacy restrictions on medical imaging data. De-identified data may be made available by the corresponding author on reasonable request for non-commercial academic research, subject to approval by the responsible ethics committee (Ethikkommission Nordwest- und Zentralschweiz) and the completion of a formal data-sharing agreement between the requesting institution and Paul Scherrer Institut. Source code supporting the methodology will be made available by the corresponding author on reasonable request.

\section*{Ethics statement}

The MR study using \SI{1.5}{\tesla} system was approved by the Ethikkommission Nordwest- und Zentralschweiz (EKNZ 2019-01060). The low-field MR study was approved by the Ethikkommission Nordwest- und Zentralschweiz (EKNZ 2023-00674). Written informed consent was obtained from all participants.


\appendix

\section{Default hyperparameter configuration}\label{app:hyperparams}

See Table \ref{tab:hyperparams}.
\begin{table}[h]
\centering
\caption{Default hyperparameter configuration used in all experiments.}
\label{tab:hyperparams}
\begin{tabular}{llc}
\toprule
Category & Parameter & Value \\
\midrule
\multirow{6}{*}{Template INR} & Architecture & Instant-NGP \\
& Hash levels $L$ & 16 \\
& Features/level $F$ & 2 \\
& Hash table size $T$ & $2^{19}$ \\
& Resolution range & 16--512 \\
& MLP hidden dims & $[64, 64]$ \\
\midrule
\multirow{3}{*}{Motion encoder} & Type & Single-frame VAE \\
& Motion code dim $d_m$ & 32 \\
& Slice-position conditioning & Sinusoidal embedding of $k/D$ \\
\midrule
\multirow{4}{*}{Deform network} & Conditioning & Adaptive \\
& Hidden layers & $[256]^4$ \\
& $\omega_0$ & 7 \\
& Bidirectional & Yes \\
\midrule
\multirow{6}{*}{Training} & Iterations & 10{,}000 \\
& Batch size & 8 slices \\
& Points/slice & 10{,}000 \\
& LR (template MLP) & $2 \times 10^{-5}$ \\
& LR (hash encoding) & $10^{-2}$ \\
& LR (encoder/deform) & $5 \times 10^{-5}$ \\
\midrule
\multirow{5}{*}{Loss weights} & $\lambda_{\text{jac}}$ & $10^{-4}$ \\
& $\lambda_{\text{mag}}$ & $0.01$ \\
& $\lambda_{\text{KL}}$ & $10^{-3}$ \\
& $\lambda_{\text{cyc}}$ & $0.1$ \\
& $\lambda_{\text{inv}}$ & $10^{-4}$ \\
\bottomrule
\end{tabular}
\end{table}

\subsection{Implementation details}

\subsubsection{Instant-NGP hash encoding formula}

Base resolution $N_{\min}=16$, maximum resolution $N_{\max}=512$ with $L=16$ levels. Per-level resolution is $N_l = \lfloor N_{\min} \cdot b^l \rfloor$ where $b = \exp(\ln(N_{\max}/N_{\min})/(L-1))$. Input coordinates are mapped to the multi-resolution grid, features at the 8 surrounding vertices are retrieved via spatial hashing and trilinearly interpolated, and all $L$ feature vectors are concatenated. The MLP decoder maps features to intensity: MLP: $\mathbb{R}^{32} \xrightarrow{\text{ReLU}} \mathbb{R}^{64} \xrightarrow{\text{ReLU}} \mathbb{R}^{64} \xrightarrow{} \mathbb{R}^{1}$ with sigmoid output.

\subsubsection{Motion encoder architecture}

ResNet-style convolutional backbone: initial $7 \times 7$ convolution (stride 2) with instance normalisation and LeakyReLU(0.2), $2 \times 2$ max pooling, six residual blocks with channel progression $32 \to 64 \to 128 \to 256$ (downsampling at channel-doubling boundaries), and adaptive average pooling to 256-dim. Sinusoidal positional embedding of slice location $k/D$ produces 32-dim encoding, refined by a single-layer projection with LeakyReLU. Concatenated features and positional embedding ($[B, 288]$) are processed by a two-layer MLP ($288 \to 144 \to d_m$) with LayerNorm, LeakyReLU, and dropout ($p = 0.1$).

\subsubsection{Gauss-Legendre quadrature nodes and weights}

5-point Gauss-Legendre quadrature on $[-1,1]$ with nodes and weights:
\begin{center}
\small
\begin{tabular}{cc}
\toprule
Node $t_s$ & Weight $w_s^{\text{GL}}$ \\
\midrule
$\pm 0.9063$ & $0.2369$ \\
$\pm 0.5385$ & $0.4786$ \\
$0.0$ & $0.5689$ \\
\bottomrule
\end{tabular}
\end{center}

\subsubsection{Training schedule and alternating evaluation mode}

Regularisation scheduling: $w_s(t) = \text{clip}(t/5, 0, 10)$ linearly ramps the DVF regularisation $\mathcal{L}_{\text{reg}}$ from zero over 50 steps, preventing early training instabilities.

Alternating evaluation-mode schedule: every 10th step, the motion encoder and deformation networks are set to evaluation mode (disabling dropout and switching instance normalisation to running statistics) while only the template INR remains in training mode. On the remaining 9 out of 10 steps, all networks are in training mode. This schedule encourages a stable canonical anatomy representation while allowing the motion networks to adapt freely.

Learning rate decay: learning rates are decayed by a factor of 0.5 every 5{,}000 steps (one decay step over the 10{,}000-iteration schedule).

\section{Extended ablation results}\label{app:extended_ablation}

\subsection{Per-case breakdown}

\begin{table*}[htbp]
\centering
\caption{Per-case reconstruction and diaphragm-landmark tracking accuracy on the \SI{1.5}{\tesla} bSSFP dataset (five volunteers $\times$ three sessions = 15 runs), proposed full configuration. Landmark MAE is computed over five CoTracker-tracked diaphragm points across the full 768-frame sequence per session.}
\label{tab:per_case}
\small
\footnotesize
\begin{tabular}{@{}llccccc@{}}
\toprule
Case & Session & PSNR & SSIM & MAE & MAE$_y$ (mm) & MAE 2D (mm) \\
\midrule
C01 & s1 & 25.80 & 0.856 & 0.029 & 5.44 & 5.94 \\
 & s2 & 25.08 & 0.836 & 0.034 & 4.97 & 5.32 \\
 & s3 & 22.37 & 0.785 & 0.049 & 5.45 & 6.03 \\
C02 & s1 & 25.05 & 0.857 & 0.031 & 2.24 & 2.92 \\
 & s2 & 26.35 & 0.889 & 0.025 & 2.03 & 2.80 \\
 & s3 & 28.96 & 0.914 & 0.021 & 1.92 & 2.50 \\
C03 & s1 & 24.49 & 0.866 & 0.033 & 2.98 & 3.26 \\
 & s2 & 27.18 & 0.898 & 0.025 & 2.69 & 2.93 \\
 & s3 & 27.08 & 0.878 & 0.027 & 3.49 & 7.77 \\
C04 & s1 & 26.78 & 0.900 & 0.025 & 3.57 & 4.23 \\
 & s2 & 30.35 & 0.932 & 0.018 & 3.74 & 4.10 \\
 & s3 & 30.07 & 0.926 & 0.019 & 3.21 & 3.45 \\
C05 & s1 & 25.30 & 0.880 & 0.027 & 2.89 & 3.99 \\
 & s2 & 26.41 & 0.883 & 0.025 & 3.31 & 4.41 \\
 & s3 & 27.41 & 0.900 & 0.023 & 2.73 & 3.83 \\
\midrule
\multicolumn{2}{l}{Mean $\pm$ Std} & 26.58$\pm$2.04 & 0.880$\pm$0.036 & 0.028$\pm$0.007 & 3.38$\pm$1.09 & 4.23$\pm$1.42 \\
\bottomrule
\end{tabular}
\end{table*}

Inter-session reproducibility is good for cases C02--C05 (PSNR standard deviation $<\!1.5$\,dB within case) and weaker for C01, whose Session 3 acquisition was visibly corrupted by large inter-cycle motion; this case disproportionately inflates the pooled standard deviation. Landmark tracking accuracy is correspondingly worse on C01 (median 2D MAE 5.76\,mm) and best on C02 (median 2.80\,mm), indicating that the method's residual error is anatomy- and acquisition-quality-limited rather than model-limited.

\bibliographystyle{elsarticle-harv}
\bibliography{media-references}

\end{document}